\documentclass[12pt]{article}
\usepackage{amsmath}
\usepackage{amsthm}
\usepackage{amssymb}
\usepackage{amsfonts}
\usepackage{epic}
\usepackage{eepic}
 \usepackage{times}
\usepackage[matrix,arrow]{xy}

\usepackage{macros}
\usepackage{color}
\usepackage{framed}

\usepackage{epsfig} 
\definecolor{shadecolor}{rgb}{1,0.9,0.7}

\theoremstyle{plain}

\newtheorem{thm}{Theorem}
\newtheorem{claim}{Claim}
\newtheorem{propn}{Proposition}

\newtheorem{conj}{Conjecture}

\newtheorem{defn}{Definition}

\theoremstyle{remark}
\newtheorem{rem}{Remark}

\newcommand{\sst}{\scriptscriptstyle}

\newcommand{\vt}{\vartheta}

\newcommand{\vac}{v_\0}

\renewcommand{\1}{\one}
\renewcommand{\2}{\two}
\newcommand{\3}{\three}
\newcommand{\4}{{\mathfrak 4}}

\newcommand{\beq}{\begin{equation}}
\newcommand{\eeq}{\end{equation}}

\newcommand{\id}{{\rm id}}

\newcommand{\pa}{\partial}
\newcommand{\ot}{\otimes}
\newcommand{\ra}{\to}

\newcommand{\ti}{\times}

\newcommand{\fsl}{{\mathfrak s}{\mathfrak l}}

\newcommand{\bra}{\langle}

\newcommand{\ket}{\rangle}

\newcommand{\df}{\equiv}

\newcommand{\al}{\alpha}

\newcommand{\be}{\beta}
\newcommand{\ga}{\gamma}
\newcommand{\Ga}{\Gamma}
\newcommand{\de}{\delta}
\newcommand{\De}{\Delta}

\newcommand{\si}{\sigma}

\newcommand{\vf}{\varphi}

\newcommand{\bw}{\bar{w}}

\newcommand{\bz}{\bar{z}}

\newcommand{\CB}{{\mathcal B}}

\newcommand{\CC}{{{\mathcal C}}}
\newcommand{\CD}{{\mathcal D}}

\newcommand{\CF}{{\mathcal F}}

\newcommand{\CH}{{\mathcal H}}

\newcommand{\CI}{{\mathcal I}}

\newcommand{\CL}{{\mathcal L}}
\newcommand{\CM}{{\mathcal M}}

\newcommand{\CT}{{\mathcal T}}

\newcommand{\CU}{{\mathcal U}}
\newcommand{\CV}{{\mathcal V}}
\newcommand{\CW}{{\mathcal W}}
\newcommand{\CX}{{\mathcal X}}
\newcommand{\CY}{{\mathcal Y}}

\newcommand{\SB}{{\mathsf B}}

\newcommand{\SE}{{\mathsf E}}
\newcommand{\SF}{{\mathsf F}}
\newcommand{\SG}{{\mathsf G}}

\newcommand{\SM}{{\mathsf M}}

\newcommand{\SP}{{\mathsf P}}  
\newcommand{\SQ}{{\mathsf Q}}  

\renewcommand{\SS}{{\mathsf S}}
\newcommand{\ST}{{\mathsf T}}
\newcommand{\SU}{{\mathsf U}}
\newcommand{\SV}{{\mathsf V}}
\newcommand{\SW}{{\mathsf W}}

\newcommand{\SY}{{\mathsf Y}}
\newcommand{\SZ}{{\mathsf Z}}

\newcommand{\fc}{{\mathfrak c}}

\newcommand{\ft}{{\mathfrak t}}

\newcommand{\sa}{{\mathsf a}}

\newcommand{\sh}{{\mathsf h}}

\newcommand{\sq}{{\mathsf q}}
\newcommand{\spp}{{\mathsf p}}

\newcommand{\0}{{\mathfrak 0}}
\newcommand{\one}{{\mathfrak 1}}
\newcommand{\two}{{\mathfrak 2}}
\newcommand{\three}{{\mathfrak 3}}

\newcommand{\BF}{{\mathbb F}}

\newcommand{\BR}{{\mathbb R}}

\newcommand{\BI}{{\mathbb I}}
\newcommand{\BC}{{\mathbb C}}

\newcommand{\BT}{{\mathbb T}}

\newcommand{\BU}{{\mathbb U}}

\newcommand{\BZ}{{\mathbb Z}}

\newcommand{\rf}[1]{(\ref{#1})}

\newcommand{\aufz}
{\begin{list}{$\bullet$}{\topsep0cm \itemsep0cm \parsep0cm}}
\newcommand{\eaufz}{\end{list}}

\begin{document}


\title{Nonrational conformal field theory}

\author{J. Teschner}

\address{DESY Theory,\\
Notkestr. 85\\
22603 Hamburg\\
Germany}

\maketitle

\section{Introduction}
\setcounter{equation}{0}

In these notes we will discuss the problem to 
develop a mathematical theory of a certain class
of conformal field theories (CFT) which contain  the 
unitary CFT.
In similar attempts of this kind the focus mostly was 
on the so-called {\it rational} CFT.
The author believes that this restriction is unnatural and
may obscure where the real issues are.
From a physical point of view it seems that
{\it rational} CFT are exceptional
rather than generic, owing their existence to some
remarkable arithmetic accidents. Although the
rational CFT are certainly 
a mathematically rich and interesting subject in its
own right, it seems to the author that 
the simplifications resulting from rationality
obscure what CFTs really are.

The present approach will be based on the so-called gluing construction
of the conformal blocks in which one constructs
large classes of conformal blocks from the conformal blocks
associated to the three punctured sphere.
Some aspects of the resulting ``Lego-Teichm\"uller game'' 
are well-understood in the case of rational CFT 
including relations to modular tensor categories, modular functors
etc., see \cite{BK1} and references therein. 
However, it seems to the author that the gluing construction had 
not yet been developed for the case of arbitrary Riemann surfaces before. 
This may be due to the fact 
that key mathematical results concerning Riemann surface theory like 
\cite{RS} have become available only recently.
The author was also unable to find a satisfactory treatment 
of the consequences of projectiveness of the canonical connection
on spaces of conformal blocks within this framework. 
We will outline an approach to CFT based on the gluing construction that
properly deals with these issues.

Of particular importance for us will be to find a 
proper generalization of the concept of a modular functor
which does not assume finite-dimensionality of the spaces of 
conformal blocks. This immediately raises the issue to
control convergence of expansions w.r.t. to a basis for (sub-) spaces
of the space of conformal blocks by means of suitable topologies.
Also for other reasons it will be seen to be of foundational 
importance to have a nondegenerate hermitian form,
or, in good cases, a scalar product on the spaces of 
conformal blocks. This is not only required for
the construction of correlation functions out of the conformal
blocks, it also serves the task to select a subspace 
of ``tempered'' conformal blocks among the space of 
all solutions to the conformal Ward identities. 
This is one of the main issues which makes the 
nonrational case much more subtle and interesting
than the rational case: As we will illustrate by an
example one will generically find that the space of 
tempered conformal blocks is much smaller than the 
space of all solutions to the conformal Ward identities.
However, the latter contains a subspace
of ``factorizable'' conformal blocks -- those that have a 
reasonable behavior at all boundaries of the moduli space 
${\mathfrak M}(\Sigma)$ of complex structures on a given two-dimensional 
surface $\Sigma$. In the example discussed
below it turns out that the space of all factorizable conformal blocks
can be fully understood\footnote{Via meromorphic 
continuation.}
provided one understands the much smaller space of all 
tempered conformal blocks. 
 
The variant of the concept 
of a modular functor that will be proposed below is based
on the consideration of stable surfaces\footnote{Surfaces $X$ with $2g-2+n>0$,
with $g$ being the genus of $X$ and $n$ the number of marked points.}
only. This is not usually done in the context of modular functors
related to rational CFT, where cutting the surface into pieces containing
discs etc. is also allowed. One of the issues that arise
is to properly formulate the distinguished role played 
by insertions of the vacuum representation. This turns out 
to be somewhat more subtle in nonrational cases.

An important issue  
is the existence of a canonical nondegenerate
hermitian form on spaces of conformal
blocks. We will propose a generalization of
known relations between the canonical hermitian form and other data 
characterizing modular functors like the so-called fusion transformation
in Section 6. 
Existence of a scalar product on 
spaces of conformal blocks seems to be an open question even for many 
rational CFT. In Section 6 
we will present arguments indicating that the hermitian form gives
a scalar product whenever one restricts attention to the 
conformal blocks associated to {\it unitary} representations.

It may also be worth mentioning the analogies between CFT
and the theory of automorphic forms \cite{Wi,Fr1,Fr2}
in which, very roughly speaking,
the role of the automorphic forms is taken 
by the conformal blocks. These analogies play an important
role in certain approaches
to the geometric Langlands-correspondence, see
\cite{Fr2} for a review. An ingredient of the classical theory
of automorphic forms that does not seem to have a good counterpart within CFT 
at the moment is a good analog of the scalar product
on spaces of automorphic forms. This structure is 
the foundation for doing harmonic analysis on 
spaces of automorphic forms.
The author believes that the scalar products on
spaces of conformal blocks discussed in this paper
provide a natural analog of such a structure.

In any case, one of my aims in this paper 
will be to advertise the
harmonic analysis on spaces of conformal blocks
as an attractive future field of mathematical research,
naturally generalizing the theory of automorphic forms
and the harmonic analysis on real reductive groups.

{\bf Acknowledgements}
The author would like to thank V. Schomerus and especially I. Runkel
for comments on a previous version of the manuscript and useful
remarks. 

Financial support from the EU by the Marie Curie Excellence
Grant MEXT-CT-2006-042695 is gratefully acknowledged.

\section{Constraints from conformal symmetry}
\setcounter{equation}{0}

\subsection{Motivation: Chiral factorization of physical correlation functions}

A point of view shared by many physicists 
is that a conformal field theory is characterized by the
set of its n-point correlation functions
\begin{equation}\label{corrfkt}
\big\langle\,V_n(z_n,\bar{z}_n)\dots V_1(z_1,\bar{z}_1)\,\big\rangle_{X}\,,
\end{equation}
which can 
be associated to any Riemann surface $X$ with $n$ marked points 
$z_1,\dots,z_n$ and a collection of vertex operators $V_k(z_k,\bar{z}_k)$
$k=1,\dots,n$. The vertex operators $V_k(z_k,\bar{z}_k)$ are in one-to-one
correspondence with states $V_k$ in representations 
${\mathfrak R}_k$ of the
conformal symmetry ${\rm Vir}\times{\rm Vir}$ 
by the state-operator correspondence. 

A lot of work on CFT was stimulated by the observation that
conformal symmetry combined with physical consistency requirements
constrain the correlation functions of a CFT strongly.

We will assume that the representations ${\mathfrak R}_k$
factorize as ${\mathfrak R}_k=R_k^{}\ot R_k'$.
It is then sufficient to know the correlation functions in the
case that the vectors $V_k\in R_k$ factorize as 
$V_k^{}=v_k^{}\ot v_k'\in R_k^{}\ot R_k'$.  The
notation  $\hat\Sigma$ will be used
as a short-hand for the topological 
surface $\Sigma$ with marked points $z_k$ ``decorated'' by the
representations $R_k$, $R_k'$.
There are general arguments which indicate that the correlation
functions should have a holomorphically factorized structure
\begin{equation}\label{holofact}
\big\langle\,V_n( z_n,\bar{z}_n) \dots 
V_1(z_1,\bar{z}_1)\,\big\rangle_{X}\,=\,
\int\limits_{\BF_{\hat\Sigma}\times \BF_{\hat\Sigma}}\!\!
d\mu_{\hat\Sigma}(S,S')\;
\CF_{S}^{}(v;X)\,\overline{\CF}_{{S}'}^{}(v';X)\,.
\end{equation}
This decomposition disentangles the relevant dependencies by 
encoding them into the
following objects:
\begin{itemize}
\item
The conformal blocks
$\CF_{S}^{}(v;X)$ and $\bar\CF_{{S}'}^{}(v';X)$ depend 
holomorphically and antiholomorphically
on the complex structure of the Riemann surface $X$, 
respectively. The set $\BF_{\hat\Sigma}$ of labels ${S}$ that 
the integration is extended
over will be specified more explicitly below. 
They furthermore depend on the vectors 
$v=\bigotimes_{k=1}^nv_k\in\bigotimes_{k=1}^n R_k$ 
and $v'=\bigotimes_{k=1}^nv_k'\in\bigotimes_{k=1}^nR_k' $, respectively.
\item 
The measure $d\mu_{\hat\Sigma}(S,S')$ does not depend on the
complex structure of the Riemann surface $X$ but only
on its topological type $\Sigma$. together with the assignment
of representations $R_k$, $R_k'$ to the punctures $z_k$.
\end{itemize}

Given that the correlation functions of a CFT factorize
as in \rf{holofact}, it has turned out to be fruitful
to approach the construction of correlation function
in three steps:
\begin{itemize}
\item First construct the conformal blocks $\CF_{S}^{}(v;X)$
by exploiting the constraints coming from the
conformal symmetry of the theory.
\item Describe the restrictions on the
measure $d\mu_{\hat\Sigma}(S,S')$ that follow from
basic physical consistency requirements
(locality, crossing symmetry, modular invartiance).
\item Identify the solution to these requirements which 
fulfils further {\it model-specific} conditions.
\end{itemize}

We will in the following mainly focus on the first two of these items.
Concerning the third let us only remark that the 
specification of the chiral symmetries will in general not 
be sufficient to determine the CFT. One may think e.g. of 
the CFTs with N=2 superconformal symmetry where one expects to 
find multi-parametric families of such CFTs 
in general.

\subsection{Vertex algebras}

Vertex algebras $V$ represent the chiral symmetries of a CFT.
We will require that
these symmetries form an extension of the Virasoro algebra
\begin{equation}\label{Vir}
[L_n,L_m]\,=\,(n-m)L_{n+m}+\frac{c}{12}n(n^2-1)\de_{n+m,0}\,.
\end{equation}
A convenient formalism for describing extensions of the 
conformal symmetry generated by the Virasoro algebra is provided
by the formalism of vertex algebras, see \cite{B,FLM,K,FBZ}.
The symmetries are generated from the modes of the 
``currents'' denoted $Y(A,z)$, with formal Laurent-expansion
of the form
\begin{equation}\label{Yexp}
Y(A,z)\,=\,\sum_{n\in\BZ} A_n z^{-n-1}\,,
\end{equation}
There is a canonical Lie algebra $U'(V)$ which can be attached to 
a vertex algebra $V$, see \cite[Section 4.1]{FBZ}. 
$U'(V)$ is generated from the expansion
coefficients $A_n$ introduced in \rf{Yexp}.

\subsection{Representations of vertex algebras}

As indicated above, one wants to assign
representations of the the vertex algebra $V$ to the 
marked points of $\Sigma$.
Representations $M$ of the vertex algebra $V$ must in particular 
be representations of the Lie-algebra $U'(V)$ generated from
the coefficients $A_n$, see \cite[Section 5]{FBZ} for more
details. 

Note that $V$ can be considered as a representation of itself, 
the so-called vacuum representation which is 
generated from a distinguished vector $\vac$ such that
$Y(\vac,z)={\rm id}$. 
This realizes the idea of state-operator correspondence: The
currents $Y(A,z)$ are in one-to-one correspondence with the
states $A\equiv A_{-1}\vac$ that they generate from the
``vacuum'' $\vac$ via $\lim_{z\ra 0} Y(A,z)\vac$.
The energy-momentum tensor $T(z)=\sum_{n\in\BZ} L_n z^{-n-2}$ is 
identified with $Y(A,z)$ for $A=L_{-2}\vac$.

We will mainly be interested in unitary representations $M$
of the vertex algebra $V$. To define the notion of a unitary
representation of $V$ we need to say how hermitian conjugation
should act on the generators of $U'(V)$. Formally this means
that we need to assume that $V$ is equipped with a
$*$-structure, a conjugate linear anti-automorphism $*:A_n\ra A_n^*$
of $U'(V)$ such that $*^2={\rm id}$.
$M$ is a unitary representation
of $V$ if it has the structure of a Hilbert space with 
scalar product $\langle.,.\rangle_M$
such that $A_n^{\dagger}=A_n^*$. We will assume in particular
that $L_n^*=L_{-n}$, as it is usually done to define unitary 
representations of the Virasoro algebra.

Let ${\BU}$ be the set of all (equivalence classes)  
of {\it irreducible} unitary representations with positive energy of $V$.  
Following the terminology from Lie group theory we will call 
${\BU}$ the unitary dual of $V$. General unitary
representations $M$ can then be parametrized by measures
$\mu_M$ on ${\BU}$. The corresponding Hilbert space
${\mathfrak H}_M$ consists of all families of vectors 
$v=(\,v_u\,;\,u\in{\BU}\,)$
such that $v_u\in M_u$ for $\mu_M$-almost 
all $u\in{\BU}$ which are 
square-integrable w.r.t. 
\begin{equation}
\lVert \,v\,\rVert^2_{}\,=\,\int_{\BU}d\mu_M(u)\;
\lVert \,v_u\,\rVert^2_{M_u}\,.
\end{equation}

We will in the following sometimes restrict attention to the 
example of the Virasoro 
algebra itself. This may be motivated by the observation that
in physics the presence of a large symmetry is a lucky accident 
rather than generic, but most of our discussion can be generalized to
other vertex algebras as well.

\subsection{Conformal blocks}\label{Confbl}

We will start by recalling a definition of the conformal blocks that
has become standard in the mathematical literature.

\subsubsection{Definition}

Let $X$ be a Riemann surface of genus $g$ with $n$ marked points 
$P_1,\dots,P_n$ and choices of local coordinates $t_k$ 
near $P_k$, $k=1,\dots,n$ such that the value $t_k=0$ parametrizes
the point $P_k\in X$.

\begin{defn}\label{CFB}
A Virasoro conformal block is a linear functional $F:
M_X\equiv\prod_{r=1}^n M_r\ra \BC$
that satisfies the following invariance condition:
\begin{equation}\label{CWI}
F\big(\,T[\eta]\cdot v\,\big)\,=\,0\,,
\end{equation}
for all $v\in M_X$ and all meromorphic vectors fields $\eta$ on $X$ that
have poles only at $z_1,\dots,z_n$. The operator $T[\eta]$ is defined
as 
\begin{equation}\label{Tdef}
T[\eta]\,=\,\sum_{k=1}^n\sum_{n\in\BZ}\eta_n^{(k)}L_n^{(k)}\,,\qquad
L_n^{(k)}={\rm id}\ot\dots\ot \underset{\rm (k-th)}{L_n}\ot
\dots\ot{\rm id}\,,
\end{equation}
where the $\eta_n^{(k)}$ are the Laurent 
expansion coefficients of $\eta$
near $P_k$,
$
\eta(t_{k})=\sum_{n\in\BZ}\,\eta_n^{(k)}\,t_{k}^{n+1}
$.
\end{defn}

The definition of conformal blocks for a general conformal 
vertex algebra $V$ is given
in \cite{FBZ}. 
In addition to the conformal invariance condition
formulated in Definition \ref{CFB} one imposes conditions which 
express invariance w.r.t. the symmetries generated by the other 
currents that generate the vertex algebra $V$.

We then denote by $\CC_V(X,R)$ the space of all
conformal blocks associated to a vertex algebra $V$,
a Riemann surface $X$ and the assignment $R$ 
of a representation
$M_k$ to each of the marked points $z_k$ on $X$.

\subsubsection{Insertions of the vacuum representation}

Let us consider the case that one of the marked points $z_0,\dots,z_n$
is decorated by the vacuum representation $V$. If e.g. $R_0=V$ we may 
compare the space $\CC_V(X,R)$ to the space $\CC_V(X',R')$
where $X'$ is  the Riemann surface obtained from $X$ by
``filling'' the marked point $z_0$, and with 
representations $R_k\in {\rm Rep}(V)$, $k=1,\dots,n$
assigned to the marked points 
$z_1,\dots,z_n$, respectively. 
It can then be shown that the spaces $\CF(X,R)$ and
$\CF(X',R')$ are canonically isomorphic
\cite[Theorem 10.3.1]{FBZ}. The isomorphism is defined by 
demanding that
\begin{equation}
F'(v)\,=\,F(\vac\ot v)\,,
\end{equation}
holds for all $v\in\bigotimes_{k=1}^n R_k$.
In other words: Insertions
of the vacuum do not change the space of conformal blocks.
This innocent looking fact will be referred to 
as the ``propagation of vacua''. It has important consequences.

\subsubsection{Deformations of the complex structure of $X$}

A key point that needs to be understood about spaces of conformal
blocks is the dependence on the complex structure of $X$. 
There is a canonical way to represent infinitesimal variations
of the complex structure on the spaces of conformal 
blocks. 
By combining the definition of 
conformal blocks with 
the so-called ``Virasoro uniformization''
of the moduli space ${\mathfrak M}_{g,n}$ 
one may construct
a representation of 
infinitesimal motions on ${\mathfrak M}_{g,n}$ on 
the space of conformal blocks.

The ``Virasoro uniformization'' 
of the moduli space ${\mathfrak M}_{g,n}$  may be formulated as 
the statement that the tangent space $T{\mathfrak M}_{g,n}$ to
${\mathfrak M}_{g,n}$ at $X$ can be identified with the
double quotient
\begin{equation}\label{VirUni}
T{\mathfrak M}_{g,n}\;=\;\Ga\big(X\setminus\{x_1,\dots,x_n\},\Theta_X\big) \;
\bigg\backslash \;
\bigoplus_{k=1}^n \BC(\!(t_k)\!)\pa_k \,
\bigg/\, \bigoplus_{k=1}^n \BC[\![t_k]\!]\pa_k\,,
\end{equation}
where $\Ga(X\setminus\{x_1,\dots,x_n\},\Theta_X)$ is the set of vector
fields that are holomorphic on $X\setminus\{x_1,\dots,x_n\}$,
while $\BC(\!(t_k)\!)$ and $\BC[\![t_k]\!]$ 
are formal Laurent and Taylor series
respectively.

Let us then consider $F(T[\eta]\cdot v)$ with 
$T[\eta]$ being defined in \rf{Tdef} in the case that
$\eta\in \bigoplus_{k=1}^n \BC(\!(t_k)\!)\pa_k$ and $L_r v_k=0$ for
all $r>0$ and $k=1,\dots,n$. 
The defining invariance property \rf{CWI} together
with $L_r v_k=0$ allow us to define
\begin{equation}\label{Viract}
\de_\vartheta F(v)=F(T[\eta_\vartheta^{}]\cdot v)\,,
\end{equation}
where $\de_{\vartheta}$ is the derivative corresponding 
a tangent 
vector $\vartheta\in T{\mathfrak M}_{g,n}$ and $\eta_\vartheta^{}$ 
is any element of $\bigoplus_{k=1}^n \BC(\!(t_k)\!)\pa_k$
which represents $\vartheta$ via \rf{VirUni}.
Generalizing these observations one is led to the
conclusion that derivatives w.r.t. to the moduli parameters
of ${\mathfrak M}_{g,n}$ are (projectively) represented on the
space of conformal blocks, the central extension coming from the 
central extension of the Virasoro algebra \rf{Vir}.

It is natural to ask if
the infinitesimal motions on ${\mathfrak M}_{g,n}$ defined above
can be integrated.
The space of conformal blocks would then have the structure of 
a holomorphic vector bundle with a projectively flat 
connection\footnote{Projective 
flatness means flatness up to a central element.}. 
This 
would 
in particular imply that locally on ${\mathfrak M}_{g,n}$ one may
define families of  conformal blocks $X\ra F_X$  
such that the
functions $X\ra F_X(v)$ depend 
holomorphically  on the
complex structure $\mu$ on $X$. 

Examples where this property has been established in full generality 
are somewhat rare, they include the WZNW-models, the minimal models 
and certain classes of rational conformal field theories in genus zero.
However, from a physicists point of view, a 
vertex algebra whose conformal blocks do not have this
property is pathological. We are not going to assume integrability
of the canonical connection in the following.

\subsection{Correlation functions vs. hermitian forms}\label{Corr-herm}

Let us return to our original problem, the problem to
construct correlation functions 
$\langle V_n(z_n,\bar{z}_n)\dots V_1(z_1,\bar{z}_1)\rangle_{X}$. 
Assuming a holomorphically factorized structure as in 
\rf{holofact}, it seems natural to identify
$\BF_{\hat\Sigma}$ with an index set for a ``basis''\footnote{Possibly in 
the sense of generalized functions.} 
$\{\CF_S(v;X);S\in\BF_{\hat\Sigma}\}$ 
for a subspace ${\mathfrak F}(\hat\Sigma)$ of the
space of solutions to the conformal Ward identities \rf{CWI} that
is defined as follows.

Let us focus attention on the dependence of $\CF_S(v;X)$ w.r.t. the label
$S$ by using the notation $f_{v,X}(S)\equiv \CF_S(v;X)$. 
The measure $d\mu_{\hat\Sigma}$
on $\BF_{\hat\Sigma}\times\BF_{\hat\Sigma}$ introduced in 
\rf{holofact} allows one to consider the space ${\mathfrak F}(\hat\Sigma)$
of functions on $\BF_{\hat\Sigma}$ such that
\begin{equation}
\int\limits_{\BF_{\hat\Sigma}\times \BF_{\hat\Sigma}}\!\!d\mu_{\hat\Sigma}(S,S')\;
(f(S))^* f(S')\;<\;\infty\,.
\end{equation}
By definition, the space ${\mathfrak F}(\hat\Sigma)$ comes equipped
with a hermitian form $H_{\hat\Sigma}$ which allows one 
to represent the correlation functions in the form
\begin{equation}\label{corrfct-herm}
\big\langle V_n(z_n,\bar{z}_n)\dots V_1(z_1,\bar{z}_1)\big\rangle_{X}\,=\,
H_{\hat\Sigma}(f_{v',X},f_{v,X})\,.
\end{equation}
The elements
of the space 
${\mathfrak F}(\hat\Sigma)$ are identified with 
elements of a subspace of the space of
conformal blocks by associating to each 
$f\in{\mathfrak F}(\hat\Sigma)$ a solution $\CF_f$ to the conformal 
Ward identities  via
\begin{equation}
\CF_f(v;X)\,\equiv\,
\int\limits_{\BF_{\hat\Sigma}\times 
\BF_{\hat\Sigma}}\!\!d\mu_{\hat\Sigma}(S,S')\;
(f(S))^*\,\CF_{S'}(v;X)\,.
\end{equation}
We are therefore confronted with the task to construct
suitable hermitian forms on subspaces of the space of 
conformal blocks which allow us to represent the correlation
functions in the form \rf{corrfct-herm}.

\section{Behavior near the boundary of moduli space}
\setcounter{equation}{0}

It is of particular importance for most applications
of CFT within physics to understand the behavior of correlation 
functions near the boundaries of the moduli space 
${\mathfrak M}(\Sigma)$ of complex structures on a 
given two-dimensional surface $\Sigma$. Such boundaries may be
represented by surfaces on which a closed geodesic $c$
was shrunk to zero length, thereby pinching a node. Two cases
may arise:
\begin{itemize}
\item[(A)] Cutting $X$ along $c$ produces two disconnected surfaces
$X_\1$ and $X_\2$ with boundary.
\item[(B)] Cutting $X$ along $c$ produces a connected surface $X'$ 
with boundary whose genus is smaller than the  genus of $X$
(``pinching a handle'').
\end{itemize}

In the following we will propose certain assumptions
which ensure existence of an
interpretation  of the CFT in question as a
quantum field theory with Hilbert space
\begin{equation}\label{Hdecomp}
{\mathfrak H}_{\rm\sst CFT}\,=\,\int_{\BU^2}^{\oplus}d\mu(r,r')
\;R_r^{}\ot R_{r'}^{}\,.
\end{equation}
These assumptions may be loosely formulated as follows.

In case (A) it is required that there exists a
representations of the correlation functions
by ``inserting complete sets of intermediate states'', schematically
\begin{equation}\label{fact1}
\big\langle\,V_n(z_n,\bar{z}_n)  \dots V_1(z_1,\bar{z}_1)\,\big\rangle_{X}
\,=\,\sum_{v\in\CB} \;
\big\langle \,0\,|\,{\mathsf O}_{X_\1}
\,q^{L_0}\bar{q}^{\bar{L}_0}\,|\,v\,\big\rangle_{X_\1}
\big\langle \,v\,|\,{\mathsf O}_{X_\2}\,|\,0\,\big\rangle_{X_\2}\,,
\end{equation}
where ${\mathsf O}_{X_r}: {\mathfrak H}_{\rm\sst CFT}^{}\ra 
{\mathfrak H}_{\rm\sst CFT}^{}$, $r=1,2$ are certain operators 
associated to the surfaces $X_\1$ and $X_\2$, respectively, and
the summation is extended over the vectors $v$ which form 
a basis $\CB$ for the space of states ${\mathfrak H}_{\rm\sst CFT}$ 
of the conformal field theory in 
question. 

In case (B) it is required that 
there exists a representations of the correlation functions
as a trace 
\begin{equation}
\big\langle\,V_n(z_n,\bar{z}_n)  \dots V_1(z_1,\bar{z}_1)\,\big\rangle_{X}
\,=\,{\rm tr}_{{\mathfrak H}_{\rm\sst CFT}}^{}
\big(q^{L_0}\bar{q}^{\bar{L}_0}{\mathsf O}_{X'}\big)\,,
\end{equation}
where ${\mathsf O}_{X'}: {\mathfrak H}_{\rm\sst CFT}^{}\ra 
{\mathfrak H}_{\rm\sst CFT}^{}$ is a certain operator 
associated to the surface $X'$.

One may formulate these two conditions more precisely
by demanding that the 
conformal blocks which appear in \rf{holofact} can be 
obtained by the gluing construction that we will now describe in more
detail.

\subsection{Gluing of Riemann surfaces}

For the following it will be more convenient to consider 
Riemann surfaces whose boundary components are represented
by holes with parametrized boundaries 
rather than marked points with choices of local coordinates around them.
Conformal invariance allows to relate the two ways of representing
boundary components, see \cite{RaS} for a mathematical discussion 
of some of the issues involved. 

\subsubsection{}\label{glue1}

Let $X'$ be a possibly disconnected Riemann surface with
$2m+n$ parametrized boundaries which are labelled 
as $C_1^+,C_1^-,\dots,C_m^+,C_m^-,B_1,\dots,B_n$.
We will assume that the parametrizations of the boundaries 
$C_r^{\pm}$, $r=1,\dots,m$
extend holomorphically to give coordinates $t_r^{\pm}$ for 
annular neighborhoods $A_r^{\pm}$ of $C_r^{\pm}$ such that the 
boundaries $C_r^{\pm}$ are represented by the circles
$|t_r^{\pm}|=1$, while the coordinates of points in the 
interior of $A_r^{\pm}$ satisfy $|t_r^{\pm}|<1$.
We will furthermore assume that the annuli $A_r^{\pm}$ are mutually
non-intersecting.

We may then define a new Riemann surface $X$ by 
identifying all the points $P_\1$, $P_\2$ which satisfy 
\begin{equation}\label{glue}
t_{r}^+(P_\1)\,t_{r}^{-}(P_\2)\,=\,q_r\,,
\end{equation} 
for given complex numbers $q_r$ such that $|q|<1$ 
and all $r=1,\dots,m$. The annuli $A_r^{\pm}$ are thereby 
mapped to annuli $A_r$ embedded into the new surface $X$.

One may apply this construction to a family $X'_t$ of surfaces
of the kind above with a set of parameters collectively 
denoted $t=(t_1,\dots,t_k)$. This yields a family $X_{q,t}$
of Riemann surfaces that is labelled by the $m+k$ parameters
$q=(q_1,\dots,q_m)$ and $t$. The family of surfaces obtained
in this way contains the nodal surfaces $X_d$ which are obtained
when at least one of the $q_r$ equals zero. If
$X'_t$ is stable, i.e. if its disconnected components all have
a number $n$ of punctures larger than $2-2g$, one 
gets the nodal 
surfaces $X_d$ that represent the points
of the Deligne-Mumford compactification 
$\overline{\mathfrak M}(\Sigma)$ of the moduli space
${\mathfrak M}(\Sigma)$ of complex structures on surfaces
$\Sigma$ homotopic to $X$.

\subsubsection{}

It will be important for us to notice that there exists a universal 
family of this kind: A family $X_p$ of surfaces such that for any other
family $Y_q$ of surfaces which contains a nodal surface $Y_{q_\0}$ 
isomorphic to $X_d$
there exists a holomorphic map
$p=\vf(q)$, defined in some neighborhood of the point $q_\0$, 
such that $Y_{q}$ and $X_{\vf(q)}$ 
are isomorphic (related by a holomorphic map).

More precisely
let us consider families $\pi_{\sst \CU}:\CX\ra \CU$ of surfaces 
degenerating into
a given nodal surface $X_d$. 
This means that $\pi$ is holomorphic and that
$X_{p}\equiv\pi^{-1}({p})$ is a 
possibly degenerate Riemann surface for each point ${p}$ in
a neighborhood $\CU$ of the boundary component
$\pa\overline{\mathfrak M}_{g,n}$ containing the
nodal surface $X_d$.  
Following \cite{RS} we will call families $\pi_{\sst \CU}:\CX\ra \CU$
as above an unfolding of the degenerate surface $X_d$. The surface
$X_d$ is called the central fiber of the unfolding $\pi_{\sst \CU}^{}$. 

Let us call a family  $\pi_{\sst \CU}:\CX\ra \CU$
universal if for any other family $\pi_{\sst \CV}:\CY\ra \CV$
which has a central fiber $Y_d$ isomorphic to $X_d$,  
there exists a unique extension of the  
isomorphism $f:Y_d\ra X_d$  
to a pair of  isomorphisms (holomorphic maps)  
$(\vf,\phi)$, where $\vf:\CV\ra\CU$ and $\phi:\CY\ra\CX$
such that $\pi_{\sst\CU}\circ\phi=\vf\circ\pi_{\sst \CV}^{}$.

\begin{thm} \label{RSthm} --- {\rm [RS]} --- \\
A nodal punctured Riemann surface $X_d$ admits a universal 
unfolding if and only if it is stable, i.e. iff $n>2-2g$.
\end{thm}

It is no loss of generality to assume that the family
surfaces $X_p$ are obtained by the gluing construction, so 
$X_p\equiv X_{q,t}$. 

\subsubsection{} \label{CUdef}

It is possible to apply
the gluing construction in the cases where $X'=\coprod_{p=1}^{2g-2+n}S_p$
is the disjoint union of three-holed spheres. 
One thereby gets
families of surfaces $X_q$ parametrized only
by the gluing parameters $q=(q_1,\dots,q_{m})$ with
$m$ being given as $m=3g-3+n$.
The different possibilities to get surfaces $X$ by gluing 
three punctured spheres can be parametrized by the choice of a cut system,
i.e. a collection ${\rm c}=\{c_1,\dots,c_{3g-3+n}\}$ 
of nonintersecting simple closed curves on $X$. $X'$ is reobtained
by cutting $X$ along the curves $c_r$, $r=1,\dots,m$. 
The coordinates $q=(q_1,\dots,q_{m})$ parametrize a 
neighborhood $\CU_c$ of the point in $\overline{\mathfrak M}_{g,n}$
represented by the maximally degenerate surface $X_c$ corresponding to $q=0$.
It is known \cite{M,HV} that one may cover all
of $\overline{\mathfrak M}_{g,n}$ with the 
coordinate patches $\CU_c$
if one considers all possible cut systems $c$ on $X$.

One may similarly use the coordinates $\tau=(\tau_1,\dots,\tau_m)$
such that $q_r=e^{-\tau_r}$ as system of coordinates for 
subsets of the Teichm\"uller space ${\mathfrak T}_{g,n}$.
One should note, however, that the coordinates $\tau$
are not unambiguously determined by the cut system $c$.
Indeed, the coordinates $\tau'$ obtained by $\tau_r'=\tau_r+2\pi i k_r$,
$k_r\in\BZ$,
would equally well do the job. 

In order to resolve this
ambiguity, one may refine the cut system $c$ by introducing a marking
$\si$ of $X$, 
a three-valent graph on $X$ such that each curve 
of the cut system intersects a unique edge of $\si$ exactly
once. 
\begin{figure}[htb] 
\epsfxsize3cm
\centerline{\epsfbox{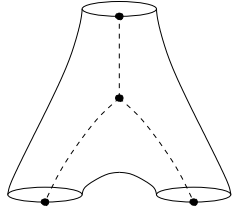}}
\caption{Standard marking of a three holed sphere.}
\label{smark}\end{figure}

With the help of the graph $\si$ one may then 
define a ``fundamental domain'' $\CV_\si$ for the variables $\tau$
as follows: Let us introduce a standard graph like the one 
depicted in Figure \ref{smark} on each of the three-holed spheres $S_p$. 
We may assume that our coordinates are such that the 
standard graphs in the three holed spheres intersect the boundaries
of the annuli
$A_r$ in the points $P^{\pm}_r$ given by
$t_r^\pm(P_r^{\pm})=|q|$, respectively. On each annulus consider 
the curve $\ga_r:[0,1]\ra A_r$, where 
$t_r^+(\ga(\theta))=|q|e^{\bar{\tau}\theta}$. The curve $\ga_r$ 
connects the points  $P^{\pm}_r$, winding 
around the annulus $A_r$ a number of times specified by 
the integer part of ${\rm Im}(\tau)/2\pi$. 
We thereby get a three valent connected graph $\si'(\tau)$ on $X$. 
The fundamental domain $\CV_\si$ 
for the coordinates $\tau$ may then be defined
by the requirement that the graph $\si'(\tau)$ is homotopic to the 
given graph $\si$.

\subsubsection{}

\begin{figure}[t]
\epsfxsize7cm
\centerline{\epsfbox{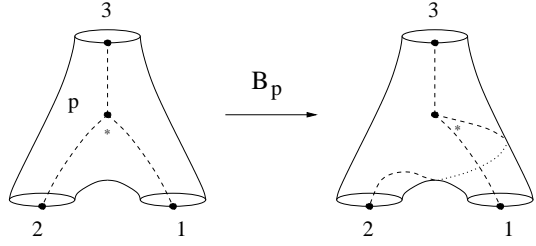}}
\caption{The B-move}\label{bmove}\vspace{.3cm}
\end{figure}

It is clear from the definitions that each marking
$\si$ determines a  cut system $c=c(\si)$. The set $\CM$ of markings
may be regarded as a cover $c:\CM\ra {\mathcal C}$
of the set ${\mathcal C}$ of all cut systems.
The subgroup ${\rm MCG}(\Sigma)_c$ of the mapping class group which preserves
a cut system $c$ acts transitively on the set of all 
markings which correspond to the same cut system. 
The group ${\rm MCG}(\Sigma)_c$ is generated by the Dehn twists 
along the curves $c_r$, $r=1,\dots,3g-3+n$ representing the cut system,
together with the braiding transformations of the three holed 
holed spheres obtained by cutting $X$ along 
the curves $c_r$, $r=1,\dots,3g-3+n$.
An example for the braiding transformations 
is graphically represented in Figure \ref{bmove}.

\subsection{Gluing of conformal blocks}

\subsubsection{}

Let us keep the set-up from paragraph \ref{glue1}. 
Let $R$ 
be an assignment of representations of a vertex algebra $V$ to the
boundary components of $X'_t$ which is such that representation
$R_r^-$ assigned to boundary component $C_r^-$ is the dual
of the representation $R_r^+$ assigned to boundary component $C_r^+$
for all $r=1,\dots,m$. Let $F_{t}^{}\in\CC_V(X'_t;R)$ be a family
of conformal blocks assigned to the family $X_t$ of surfaces with
an assignment $R$ of representations to the boundary components
of $X_t$ as above.
Let finally 
\[ e_\tau\,\equiv\,\bigotimes_{r=1}^m e_r(\tau_r)\,,\qquad
e_r(\tau_r)\,\equiv\,
\sum_{\imath\in\CI_r}{e}_{r,\imath}^-\ot 
e^{-\tau_rL_{0,r}} e_{r,\imath}^+\,,
\]
where
$\{{e}_{r,\imath}^+;\imath\in \CI_{r}\}$
and $\{{e}_{r,\imath}^-;\imath\in \CI_{r}\}$ are bases for
$R_r^+$ and $R_r^-$, respectively, which are dual to each other
in the sense that $\langle{e}_{r,\imath}^-,
e_{r,\jmath}^+\rangle_{R_r^+}
=\de_{\imath,\jmath}$, with 
$\langle.,.\rangle_{R_r^+}$ being the dual pairing.

We may then consider the expression
\begin{equation}\label{factor}
G_{t,\tau}(v)\,{=}\,
F_t(v\ot e_\tau)\,,
\end{equation} 
where $v_1\ot\dots \ot v_n$.
As it stands, the expression is defined
as a formal power series in powers of $e^{-\tau_r}$.

\subsubsection{}

To proceed, we will need to introduce a nontrivial 
assumption:
\begin{quote}
The series in \rf{factor} have a finite radius of convergence.
The resulting domains of definition $\CD_{\si}$  of the conformal blocks
$G_{t,\tau}$ 
cover
the neighborhoods $\CU_{c(\si)}\subset
\overline{\mathfrak M}_{g,n}$ which form an atlas of 
$\overline{\mathfrak M}_{g,n}$ according to \cite{M,HV}.
\end{quote}

This assumption about the analyticity of the conformal 
blocks $G_{t,\tau}$ obtained from the gluing construction
is fundamental for most of the rest. On the other hand
it is not yet clear  how large the class of vertex
algebras is for which they are satisfied.

For rational CFT it is to be expected that the 
conformal blocks satisfy a closed system of 
differential equations 
which allow one to show convergence and existence of an 
analytic continuation. 
The situation is more difficult 
in the case of nonrational CFT. Validity of the assumption 
above is known  \cite{TL} in the case of the Virasoro algebra for
surfaces with $g=0$. This leads one  to
suspect that the assumption above should be 
valid in genereral, as it should always be possible
to decompose the conformal
blocks of a vertex algebra $V$ into
Virasoro conformal blocks.  
This issue is further discussed in 
Remark \ref{convrem} below. 

\subsubsection{Conformal Ward identities}

We want to check that the expression in 
\rf{factor} satisfies the conformal Ward identities.
To this aim it suffices to notice that 
$\eta$ can be split as the sum of
$\eta_{\rm\sst out}$ and $\eta_{\rm\sst in}$,
where $\eta_{\rm\sst out}$ is holomorphic in $X'$,
and $\eta_{\sst{\rm in}}$ is holomorphic in $\bigcup_{r=1}^{3g-3+n}A_r$.
This means that $G_{t,\tau}(T[\eta] v)$ can be represented 
as
\begin{equation}
G_{t,\tau}\big(T[\eta] v\big)\,=\,
F_t\big(T[\eta_{\rm\sst out}]( v\ot 
e_\tau)\big)+F_t\big( v\ot 
T[\eta_{\rm\sst in}] e_\tau\big) \,.
\end{equation} 
We have $F_t(T[\eta_{\rm\sst out}]w)=0$ 
for all $\eta_{\rm\sst out}$ that are holomorphic in $X'_t$
by the conformal Ward identities satisfied by $F_t$.
It is furthermore possible to show that the vectors
$e_r(\tau_r)$ are invariant under the action of the holomorphic 
maps $\eta_{\sst{\rm in}}^r\equiv \eta_{\sst{\rm in}}|_{A_r}$, 
\[ T[\eta_{\rm\sst in}^r]e_r(\tau_r)\,=\,0\,,
\qquad r=1,\dots,m\,.
\]
The conformal Ward identities  follow from the combination 
of these  two observations.

Keeping in mind Theorem \ref{RSthm} from \cite{RS}
it follows from the conformal Ward identities 
that the conformal blocks defined in \rf{factor}
do not depend on the choices involved in the gluing
construction. 
The resulting families $G_{t,\tau}$ of conformal blocks 
are therefore well-defined in a neighborhood $\CV_\si$ of the 
component of $\pa{\mathfrak T}_{g,n}$ that 
is obtained by $\tau_r\ra \infty$ for $r=1,\dots,m$.

It seems to the author that the foundational importance for
CFT of the work \cite{RS} was hitherto 
not as widely appreciated as it should be.
It explains in particular why it is absolutely preferable to 
formulate CFT in terms of stable surfaces.

\subsubsection{Decorated marking graphs}

\begin{figure}[t]
\epsfxsize6cm
\centerline{\epsfbox{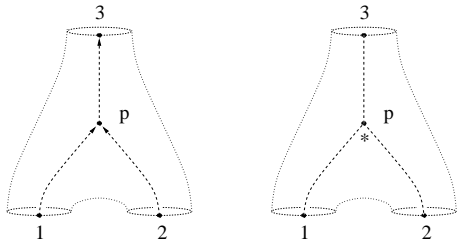}}
\caption{Two representations for the decoration on a marking graph}\label{marking}
\end{figure}

In the gluing construction above one assigns elements of a
basis for the space of conformal blocks of the three punctured 
sphere to each vertex of a marking graph. The definition of 
a basis for this space will generically depend on the choice of 
a distinguished boundary component of the three punctured sphere
in question. In order to parametrize different bases in spaces of
conformal blocks one needs to choose for each vertex $p\in\si_\0$
a distinguished edge emanating from it. The distinguished
edge will sometimes be called ``outgoing''. 
As a convenient
graphical representation we will use the one introduced 
in Figure \ref{marking}. The term marking will henceforth
be used for graphs $\si$ as defined above together 
with the choice of a distinguished
edge emanating from each vertex.

\subsubsection{}

The following data label the conformal blocks on $X$ 
that can be constructed by means of the gluing construction.
We will denote by $\si_\0$ and $\si_\1$ the sets of vertices and 
edges of $\si$, respectively.
\begin{itemize}
\item The marking $\si$.
\item An assignment $\rho$ of representation labels
$r_e\in\BU$ to the edges $e\in\si_\1$.
\item An assignment $w$ of conformal blocks $w_p\in
\CC_{V}^{}(S_p,\rho_p)$ to each 
vertex $p\in\si_\0$ of $\si$ with assignment $\rho_p$
of representations to the edges that emanate from $p$ determined by $\rho$.
\end{itemize}
We will use the notation 
$G_{\si\tau}(\de)$ for the family of conformal
blocks which is essentially uniquely defined by the
data $\si$ and the ``decoration'' $\de=(\rho,w)$.

\subsection{Correlation functions}

We are now in the position to formulate our requirements
concerning the behavior of the correlation 
functions $\langle V_n(z_n,\bar{z}_n)\dots V_1(z_1,\bar{z}_1)\rangle_{X}$
near the boundary of moduli space more precisely.
Let us consider a
marking $\si$. The marking $\si$ determines a maximally degenerate 
surface $X_{c(\si)}$ and a neighborhood $\CU_{c(\si)}
\subset\overline{\mathfrak M}_{g,n}$ of 
$X_{c(\si)}$.
Let us recall the set-up from subsection \ref{Corr-herm}, in 
particular the representation \rf{corrfct-herm}. We will 
adopt the following two requirements:

\subsubsection{First requirement: Factorization of conformal blocks}

The functions $\CF_S^{}(v;X)$ can be identified 
with the values $G_{\si\tau}(\de|v)$ of 
the conformal blocks obtained from the gluing construction
provided that a suitable identification
between the labels $S$  and the decorations $\de$ is adopted. This means
in particular 
that the space $\BF_{\hat\Sigma}$ of all labels $S$ should be
identified with the {\it sub}-space $\CI_{\si,R}$ 
of the space $\CI_\si$ of all decorations $\de$ which contains 
all decorations $\de_R$ with fixed assignment $R$ of 
representations to the external egdes\footnote{The edges
ending in the boundary components of $\Sigma$} 
of $\si$.

The space $\CI_\si$ of all 
decorations can be described more explicitly as
\begin{equation}
\CI_\si\,\equiv\,
\prod_{\rho\in\BU^{\si_\1}} \bigotimes_{p\in\si_\0}\CI_p(\rho_p) \,,
\end{equation}
where $\BU^{\si_\1}$ is the vector
space of all assignments $\rho$ of elements $r_e\in \BU$ 
to the edges $e\in\si_\1$ and  $\CI_p(\rho_p)$ is the 
index set for a basis in $\CC(S_p,\rho_p)$, $p\in\si_\0$.

The role
of the functions $f_{v,X}$ from  subsection \ref{Corr-herm} is then 
taken
by the functions
$E_{\si\tau}^R(v)$
which map a decoration $\de_R\in\CI_{\si,R}$ to
$G_{\si\tau}(\de_R|v)$. We will later find it more convenient to 
consider the functions $E_{\si\tau}(v):\CI_\si\ra \BC$
which map an {\it arbitrary} decoration $\de\in\CI_{\si}$ to
$G_{\si\tau}(\de|v)$.
This is natural in view of the fact that
the assignment $R$ of representations to boundary components
is part of the data contained in the decoration $\de$.
The space of all complex-valued functions on $\CI_\si$
will be denoted ${\mathfrak F}_\si$.

\subsubsection{Second requirement: 
Factorization of the hermitian form $H_{\hat{\Sigma}}$} \label{H-3pt}

In order to have a representation for the correlation functions in the
form of \rf{corrfct-herm} there must exist
a hermitian  form 
$H_\si^{R'R}$ on suitable 
subspaces of ${\mathfrak F}_\si\times {\mathfrak F}_\si$ 
such that
\begin{equation}\label{holofact+}
\big\langle\,V_n(z_n,\bar{z}_n)  \dots V_1(z_1,\bar{z}_1)\,\big\rangle_{X}
\,=\,
H_{\si}^{R'R}\big(\,E_{\si\tau}(v')\,,\,E_{\si\tau}(v)\,\big)\,.
\end{equation} 
Our second main requirement is that
the hermitian form $H_\si^{R'R}$ can be factorized as
\begin{equation}\label{Hfactor}
H_\si^{R'R}\,=\,\int\limits_{\BU^{\si_\1}\times\BU^{\si_\1}}
d\mu_{R'R}^{}(\rho',\rho)\;\bigotimes_{p\in\si_\0}
H_p^{\rho_p'\rho_p^{}}\,,
\end{equation}
where 
\begin{itemize}
\item the measure $d\mu_{R'R}$ has support only when the 
assignment of representations to the external edges given by $\rho'$ and $\rho$
coincides with $R'$ and $R$, respectively, and
\item
the  hermitian forms $H_p^{\rho_p'\rho_p^{}}$ 
are defined on certain subspaces of the spaces ${\mathfrak F}(\rho_p')
\ti{\mathfrak F}(\rho_p^{})$ 
of complex-valued functions on
$\CI(\rho'_p)\times \CI(\rho_p^{})$.
\end{itemize}

Assuming that the hermitian forms  $H_p^{\rho'\rho}$ 
can be represented in the form
\begin{equation}\label{Hrep}
H_p^{\rho'\rho}(f,g)\,=\,\sum_{\imath',{\imath}\in\BI} \,(f(\imath'))^*\,
H_{\imath'{\imath}}(\rho',{\rho})\,g({\imath})\,,
\end{equation}
one may in particular represent the three point functions 
$\langle V_\3(y_\3,\bar{y}_\3) V_\2(y_\2,\bar{y}_\2)   
V_\1(y_1,\bar{y}_\1)\rangle_{S_{0,3}}$ as 
\begin{align}\label{Hthree}
\big\langle\,    V_\3(\infty)    V_\2(1)  & 
V_\1(0)\,\big\rangle_{S_{0,3}}\,=\,\\
&=\,\sum_{\imath',{\imath}\in\BI}\,
H_{\imath'{\imath}}(\rho',{\rho})\;
G_{\imath'}
(\,\rho'\,|\,v_\3'\ot v_\2'\ot v_\1')_{}\;
G_{{\imath}}(\,\rho\,|\,
v_\3^{}\ot v_\2^{}\ot v_\1^{})_{}\,.
\nonumber\end{align}
In \rf{Hthree} we have used the standard model for $S_{0,3}$ 
as ${\mathbb P}^1\setminus\{0,1,\infty\}$ and assumed that 
$y_3=\infty$, $y_2=1$, $y_1=0$.
$\{G_{\imath'}(\rho');\imath'\in\CI(\rho')\}$ 
and $\{G_{\imath}(\rho);\imath\in\CI(\rho)\}$  are bases for 
$\CC(S_{\sst 0,3}.\rho')$ and $\CC(S_{\sst 0,3}.\rho)$,
respectively. 

\subsection{Conformal blocks as matrix elements}

For future use, let us note that
in  the case that $X$ has genus zero there is a convenient 
reformulation of the gluing construction of the 
conformal blocks in terms of chiral vertex operators.

\subsubsection{Chiral vertex operators} \label{CVO}

\begin{figure}[htb] 
\epsfxsize3cm
\centerline{\epsfbox{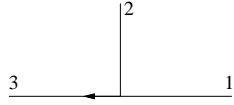}}
\caption{Diagrammatical representation for chiral vertex operators.}
\label{CVOfig} \end{figure}

The chiral vertex operators are families of 
operators $\SY_{r_\3r_\1}^{r_\2}({\mathfrak v}_\2|z)$,
${\mathfrak v}_\2\in R_\2$, $z\in\BC\setminus\{0\}$
mapping the representation $R_{r_\1}$ to the dual $\bar R_{r_\3}$
of $R_{r_\3}$. They 
are defined such that
the conformal blocks $G^{\sst (3)}$ 
associated to the three punctured
sphere $S^{\sst (3)}$ are related to the matrix elements
of $\SY_{r_\3r_\1}^{r_\2}({\mathfrak v}_\2|z)$ as
\begin{equation}\label{Ydef}
F^{\sst (3)}(\,\rho\,|\,
{\mathfrak v}_\3\ot{\mathfrak v}_\2\ot {\mathfrak v}_\1)\,=\,
\big\langle\, {\mathfrak v}_\3\,,\,
{\mathsf Y}_{r_\3r_\1}^{r_\2}({\mathfrak v}_\2|\,1\,)\,
{\mathfrak v}_\1\,\big\rangle_{R_\3}\,.
\end{equation}
It is assumed that the assignment $\rho:k\mapsto r_k$, $k\in\{1,2,3\}$
is in correspondence to the numbering of boundary components
introduced in Figure \ref{marking}. A simplified diagrammatical 
representation is introduced in 
Figure \ref{CVOfig}.

The chiral vertex operators 
$\SY_{r_\3r_\1}^{r_\2}({\mathfrak v}_\2|z)$ are
well-defined by \rf{Ydef} in the sense of formal power series in $z$.

\subsubsection{}

\begin{figure}[htb] 
\epsfxsize8cm
\centerline{\epsfbox{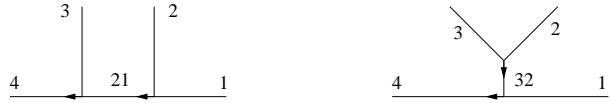}}
\caption{Diagrammatic representation for 
the compositions in \rf{compo}.}
\label{compofig}\end{figure}

There are two different ways to glue two decorated
three-punctured spheres
such that the outgoing boundary component of the first
is glued to one of the incoming
boundary components of the other. To each of the two
gluing patterns we may associate two natural ways
to compose chiral vertex operators, namely
\begin{equation}\label{compo}
{\mathsf Y}_{r_\4 r_{\2\1}}^{r_\3}(\,{\mathfrak v}_\3\,|\,1\,)
{\mathsf Y}_{r_{\2\1} r_\1}^{r_\2}(\,{\mathfrak v}_\2\,|\,z\,)
\,{\mathfrak v}_\1
\quad
{\rm and}\quad 
{\mathsf Y}_{r_\4 r_\1}^{r_{\3\2}}\big(\,
{\mathsf Y}_{r_{\3\2} r_\2}^{r_\3}(\,{\mathfrak v}_\3\,|\,1-z\,)\,
{\mathfrak v}_\2\,|\,1\,\big)
\,{\mathfrak v}_\1
\,,
\end{equation}
respectively. A diagrammatic representation for these two ways
to compose chiral vertex operators is given in Figure \ref{compofig}, 
respectively.

Let us call a marking $\si$ on a surface 
$X$ of genus 0 admissible if the 
outgoing boundary components of one pair of
pants are always
glued to an incoming 
boundary component of another.
The resulting markings $\si$ distinguish a 
unique outgoing boundary 
component of the surface $X$. We will assign the representation $R_n$
to this boundary component.
Using the  rule for compositions of chiral vertex operators
recursively we can construct an operator 
${\mathsf Y}_{X,\si}:R_{n-1}\ot \dots \ot R_\1\ra \bar R_{n}$ 
such that the conformal block 
associated to $\si$ can be represented as 
\[ 
G_{\si}({\mathfrak v}_n  \ot\dots\ot {\mathfrak v}_1)\,=\,
 \big\langle  {\mathfrak v}_{n}\, ,\, 
 {\mathsf Y}_{X,\si}({\mathfrak v}_{n-1}\ot\dots\ot
{\mathfrak v}_{1})\,  \big\rangle_{R_n}^{} \; ,
\]
for all ${\mathfrak v}_{k}\in R_k$, $k=1,\dots, n$.
The fact that the matrix elements above represent the conformal 
blocks follows from the observation that the composition of 
chiral vertex operators is equivalent to the gluing operatation.

\begin{rem} \label{convrem} The issues of
convergence of the power series defined by the gluing construction
and convergence of the power series representing the 
chiral vertex operators are closely
related. In rational CFT one can settle this issue with the help
of the differential equations satisfied by the conformal blocks,
see e.g. \cite{TK}. 
In the case of the Virasoro algebra one may use 
analytic arguments for deriving such results \cite{TL}.
The typical situation seems to be that the power series
representing, for example,
$\langle {\mathfrak v}_{n+1},
{\mathsf Y}_{r_{n+1} s_{n-1}}^{r_{n}}(\,{\mathfrak v}_{n}\,|\,z_{n}\,)
\dots 
{\mathsf Y}_{s_{\1} r_\0}^{r_\1}(\,{\mathfrak v}_\1\,|\,z_1\,)
\,{\mathfrak v}_\0\rangle
$
converge provided that the variables $z_k$ are {\it radially} ordered
$|z_n|>\dots>|z_{1}|$.
\end{rem}

\section{From one boundary to another}\label{Physreq}
\setcounter{equation}{0}

For a given pair of markings $\si_\2$, $\si_\1$ it may happen that 
the domains  $\CD_{\si_r}$, $r=1,2$ in
which the corresponding conformal blocks can be defined 
by means of the gluing construction 
have a nontrivial overlap,
$\CD_{\si_\2}\cap \CD_{\si_\1}
\neq\emptyset$. Assume that $\tau_1$ and $\tau_2$ 
parametrize the same point in 
$\CD_{\si_\2}\cap \CD_{\si_\1}\subset{\mathfrak T}_{g,n}$.
We then have two possible 
ways to represent the correlation function 
$\langle\,V_n(z_n,\bar{z}_n)  \dots V_1(z_1,\bar{z}_1)\,\rangle_{X}$
in the form \rf{holofact+}, namely either as 
$H_{\si_\1}^{}(G_{\si_\1\tau_\1},
G_{\si_\1\tau_\1})$ or as 
$H_{\si_\2}^{}(G_{\si_\2\tau_\2},G_{\si_\2\tau_\2})$, 
respectively.
It is a natural requirement to demand that these
two representations agree,
\begin{equation}\label{crossing}
H_{\si_\1}^{}\big(\,G_{\si_\1\tau_\1}\,,\,
G_{\si_\1\tau_\1}\,\big)\,=\,
H_{\si_\2}^{}\big(\,G_{\si_\2\tau_\2}\,,\,G_{\si_\2\tau_\2}\,\big)\,.
\end{equation}
These constraints  generalize what is often called
crossing symmetry, locality and modular invariance. 
Keeping in mind our assumption that the domains $\CD_{\si_r}$ cover
the neighborhoods $\CU_{c(\si)}\subset
\overline{\mathfrak M}_{g,n}$ which form an atlas of 
$\overline{\mathfrak M}_{g,n}$ 
we arrive at an unambiguous definition of the correlation functions
on all of ${\mathfrak M}_{g,n}$. 

In order to analyze the conditions further, we need to 
introduce another assumption: 
\begin{center}
The families
$G_{\si_\1\tau_\1}$ and $G_{\si_\2\tau_\2}$ are linearly related.
\end{center}
This assumption  will be formulated more precisely below. 
It is, on the one hand, 
absolutely necessary for the validity of \rf{crossing} at least
in the case of rational CFT \cite{MS}\footnote{It seems likely
that the argument in \cite{MS} can be generalized 
to the nonrational case if there exists an analytic 
continuation of the correlation functions from the euclidean section 
$\bar{\tau}=\tau^*$ to functions analytic in independent
variables $\tau,\bar{\tau}$.}. The assumption above is, on the
other hand, a 
rather deep statement about a given
vertex algebra $V$ from the mathematical 
point of view, especially if $V$ is not rational.

What simplifies the analysis somewhat is the fact that the 
relations between pairs of markings can be reduced to
a few simple cases associated to Riemann surfaces of low
genus $g=0,1$ and low number of marked points $n\leq 4$.
In order to explain how this reduction works we will
begin by briefly reviewing the necessary results from
Riemann surface  theory.

\subsection{The modular groupoid}

The physical requirements above
boil down to understanding the relations between conformal 
blocks associated to pairs $[\si_\2,\si_1]$ of markings. 
In order to break down understanding such relations to 
a sort of Lego game it will be very useful to observe
that all transitions between two markings can be factorized
into a simple set of elementary moves. One may formalize
the resulting structure by introducing a two-dimensional 
CW complex $\CM(\Sigma)$ with set of vertices $\CM_\0(\Sigma)$
given by the
markings $\si$, set of edges $\CM_\1(\Sigma)$ associated to the elementary
moves. It will then be possible
to identify the set $\CM_\2(\Sigma)$
of ``faces''  (relations between the elementary
moves) that ensure simply-connectedness of  $\CM(\Sigma)$, as we are
now going to describe in more detail.

\subsubsection{Generators}\label{MS-gens}

\begin{figure}[t]
\epsfxsize6cm
\centerline{\epsfbox{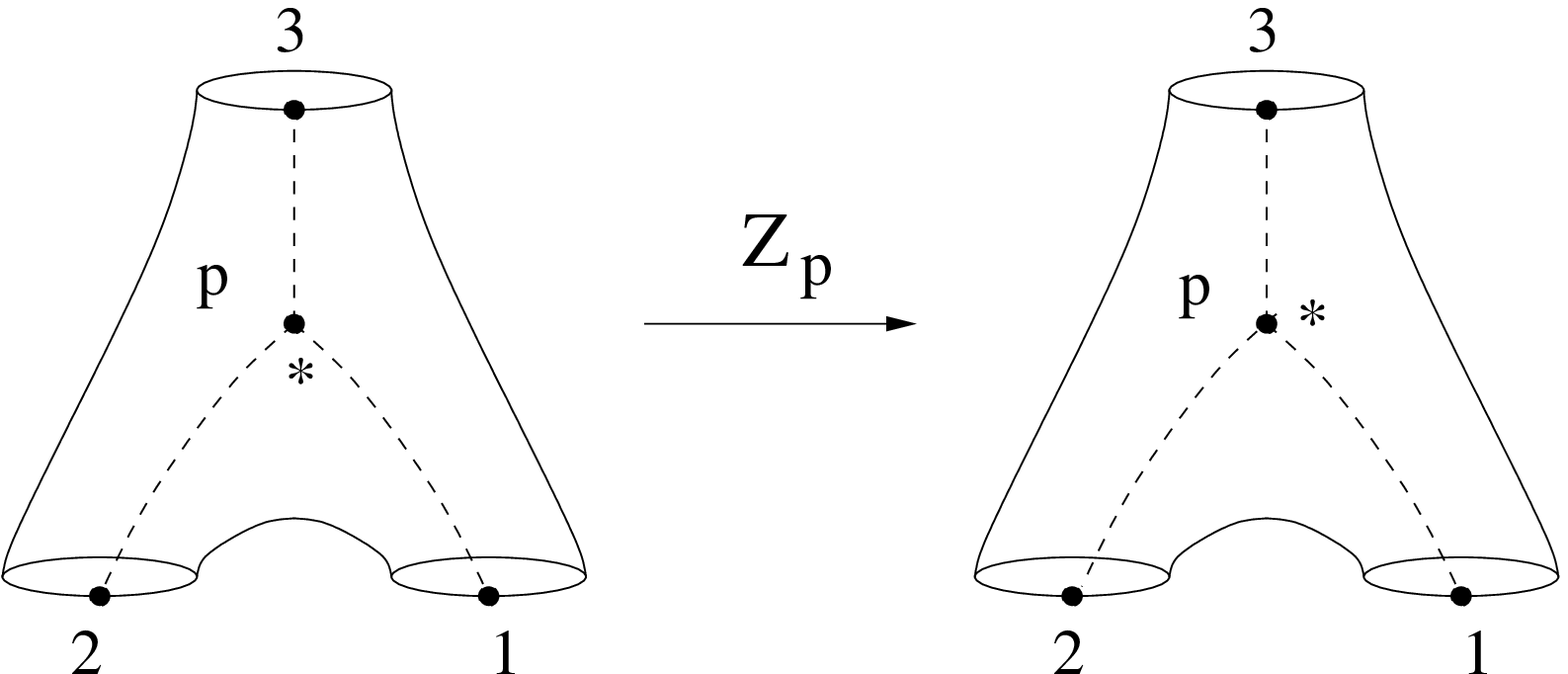}}
\caption{The Z-move}\label{zmove}\vspace{.3cm}
\epsfxsize7cm
\centerline{\epsfbox{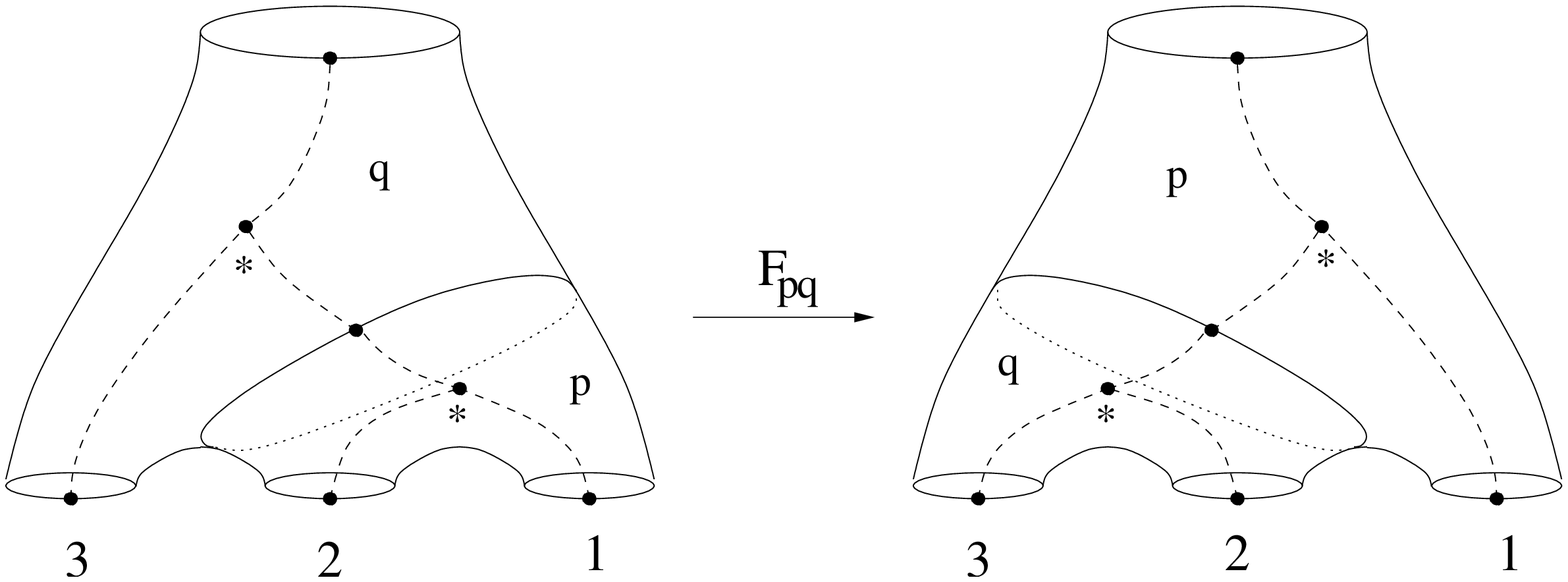}}
\caption{The F-move}\label{fmove}\vspace{.3cm}
\epsfxsize8cm
\centerline{\epsfbox{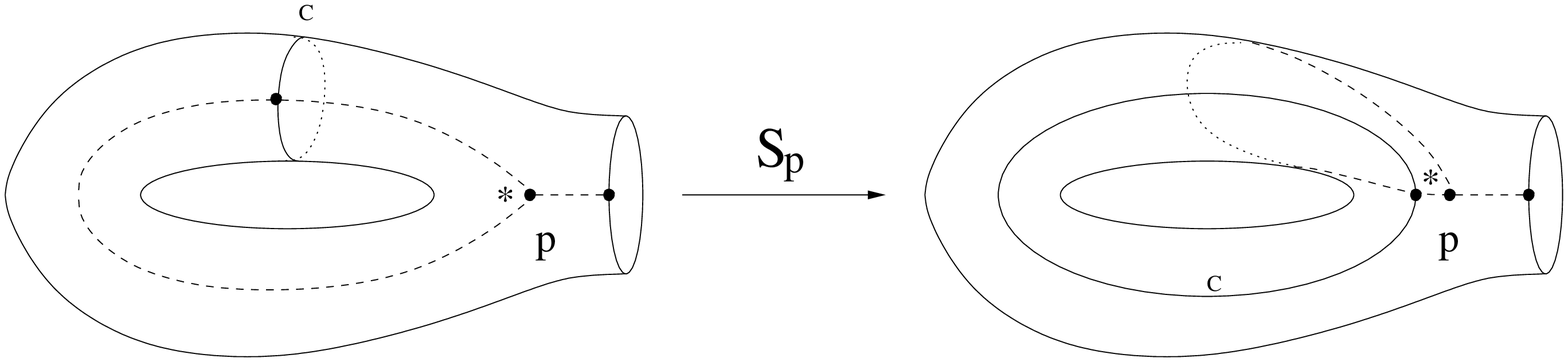}}
\caption{The S-move}\vspace{.3cm}
\label{smove}\end{figure}

The set of edges ${\CM}_{\1}(\Sigma)$ of $\CM(\Sigma)$
will be given by elementary
moves denoted as $(pq)$,
$Z_p$, $B_p$, $F_{pq}$ and $S_p$. The indices $p,q\in\si_{\0}$ 
specify the relevant three holed spheres within the 
pants decomposition of $\Sigma$ that is determined by $\sigma$. 
The move $(pq)$ will simply be the operation in which the
labels $p$ and $q$ get exchanged.
Graphical representations for the elementary 
moves $Z_p$, $B_p$, $F_{pq}$ and $S_p$ are given in 
Figures \ref{zmove}, \ref{bmove},  \ref{fmove} and \ref{smove}, respectively.


\begin{propn} {\rm \cite{BK2}} \label{connect}
Any two markings $\si$, $\si'$ can be connected to 
each other by a path composed out of the 
moves $(pq)$,
$Z_p$, $B_p$, $F_{pq}$ and $S_p$.
\end{propn}

\subsubsection{Relations}\label{rels}

It is useful to notice that {\it any} round trip can be broken
up into elementary ones. A complete list of these elementary 
round trips was first presented in \cite{MS}, see \cite{BK2}
for a mathematical treatment.

To simplify notation we will write $\pi_\2\sim\pi_\1$ if the
round trip $\pi$ in question can be represented as
$\pi=\pi_\2^{}\circ\pi_\1^{-1}$.

\paragraph{Relations supported on surfaces of genus zero.}

\newcommand{\LR}{\quad\Leftrightarrow\quad}
\begin{align}\label{zrel}
g=0,~~s=3:\quad & Z_p\circ Z_p\circ Z_p \sim \id \, .\\[1ex]
g=0,~~s=4:\quad & 
\begin{aligned}\label{hexarel} {}{\rm a)} \quad& 
F_{qp}^{\phantom{1}}\circ B_p\circ F_{pq}^{\phantom{1}}\;\sim\;
(pq)\circ B_q \circ F_{pq}^{\phantom{1}}\circ B_p\,,\\
{\rm b)}\quad & F_{qp}^{\phantom{1}}
\circ B_p^{-1}\circ F_{pq}^{\phantom{1}}\;\sim\;
(pq)\circ B_q^{-1} \circ F_{pq}^{\phantom{1}}\circ B_p^{-1},\\
{\rm c)} \quad & A_{pq}\circ A_{qp}\;\sim\;(pq)\,.
\end{aligned}\\[1ex]
g=0,~~s=5:\quad & F_{qr}\circ F_{pr}\circ F_{pq}\;\sim\; F_{pq}\circ F_{qr}.
\label{pentarel}\end{align}
We have used the abbreviation
\begin{equation}\label{Adef}
A_{pq}\;\df\; Z_q^{-1}\circ
F_{pq}^{\phantom{1}}\circ Z_q^{-1}\circ Z_p^{\phantom{1}}\,.
\end{equation}
\paragraph{Relations supported on surfaces of genus one.}

In order to write the relations transparently let us introduce the
following composites of the elementary moves. 
\begin{align}
\label{b-composites}g=0,~~s=3:\qquad & \begin{aligned}
{\rm a)}\quad & 
B_p'\;\df\; Z_p^{-1}\circ B_p\circ Z_p^{-1},\\
{\rm b)}\quad & T_{p} \;\df\; Z_p^{-{1}}\circ
B_p\circ Z_p^{\phantom{1}}\circ B_p\, ,
\end{aligned}\\[1ex] \label{bcompdef}
g=0,~~s=4:\qquad &
B_{qp}\,\df\, Z_q^{-1}\circ F_{qp}^{-1}\circ B_q'
\circ F_{pq}^{-1}\circ Z_q^{-1}\circ(pq)
\,,\\[1ex]
g=1,~~s=2:\qquad &
S_{qp}\;\df\; (F_{qp}\circ Z_q)^{-1}\circ S_p\circ (F_{qp}\circ Z_q)\,.
\label{scompdef}\end{align}
It is useful to observe that the move
$T_{p}$,  represents the Dehn twist around the 
boundary component of the trinion $\ft_p$ numbered by $i=\1$ 
in Figure \ref{marking}.

With the help of these definitions we may write the 
relations supported on surfaces of genus one as follows:
\begin{align}\label{onetor:a}g=1,~~s=1:\quad & 
\begin{aligned}{\rm a)}\quad &
S^2_p \;\sim\; B'_p,\\
{\rm b)}\quad & S_p^{\phantom{1}}\circ  T_p^{\phantom{1}}
\circ S_p^{\phantom{1}}\;\sim\;T^{-1}_p\circ S_p^{\phantom{1}}\circ T^{-1}_p .
\end{aligned}\\[1ex]
g=1,~~s=2:\quad & 
B_{qp}^{\phantom{1}}\;\sim\;S_{qp}^{-1}\circ T_{q}^{-1}
T_{p}^{\phantom{1}}
\circ S_{pq}^{\phantom{1}}\,.
\label{twotorrel}\end{align}

\begin{thm}\label{BKthm} -- {\rm \cite{BK2}} --
The complex $\CM(\Sigma)$ is connected and simply connected for 
any e-surface $\Sigma$.
\end{thm}

\subsection{Representation of the generators on spaces of conformal blocks}

\begin{defn}
We will say that a vertex algebra $V$ has the factorization
property if 
\begin{itemize}
\item[(i)] 
there exists an analytic 
continuation of the family $G_{\si\tau}({\rho,w})$ 
into a domain in ${\mathfrak T}_{g,n}$ 
that contains the fundamental domain $\CV_{\si}$ 
in ${\mathfrak T}_{g,n}$. 
\item[(ii)] 
For each pair $(\si_\2,\si_\1)$ of markings related by one of the 
elementary transformations $Z_p$, $B_P$, $F_{pq}$ and $S_p$
there exists a nontrivial intersection 
$\CD_{\si_\2\si_\1}\subset{\mathfrak T}_{g,n}$ of the 
domains $\CD_{\si_\2}$ and $\CD_{\si_\1}$ within which the 
conformal blocks $G_{\si_\1\tau_\1}(\rho_\1,w_\1)$  and
$G_{\si_\2\tau_\2}^{}(\rho_\2,w_\2)$ can be uniquely 
defined by the gluing 
construction and analytic continuation. 
There exists a relation of the form
\begin{equation}\label{genfusion}
G_{\si_\1\tau_\1}
(\rho_\1,w_\1)\,=\,\int d\mu_{\sst\Sigma}^{}(\rho_\2)\,
\sum_{w_\2} \,F_{\si_\1\si_\2}(\rho_\1|\rho_2)_{w_\1}^{w_\2}\,
G_{\si_\2\tau_\2}^{}(\rho_\2,w_\2)\,,
\end{equation}
which holds whenever $\tau_\2$, $\tau_\1$ parameterize
the same point in $\CD_{\si_\2\si_\1}$.
\end{itemize}
\end{defn}

Let us note that the conjecture can be verified
by elementary means in the cases of the moves $Z_p$ and $B_p$.
These are the moves which do not change the cut system.
Existence of a relation like \rf{genfusion} for the cases that 
$\si_\2$ and $\si_\1$ are related by a $F_{pq}$- or $S_p$-move
is a deep statement. It is, on the one hand, a 
requirement without which a CFT can hardly be of 
physical relevance. It may, on the other hand, be hard to prove
mathematically that a given vertex operator algebra has this
property.
Statements of this type
are presently only known in the case of certain rational 
vertex algebras (from the differential equations satisfied by the conformal
blocks) 
or in the nonrational case of the Virasoro algebra reviewed in 
Section \ref{Liourev}.

\subsubsection{}

The relation \rf{genfusion} suggests to consider 
an operator ${\mathsf M}_{\si_\1\si_\2}$ between certain subspaces of
${\mathfrak F}_{\si_\2}$ and ${\mathfrak F}_{\si_\1}$, respectively,
which is defined such that
\begin{equation}\label{Fdef}
{(\mathsf M}_{\si_\2\si_\1} f)_{w_\1}^{}(\rho_\1)
\,=\,\int d\mu_{\sst\Sigma}^{}(\rho_\2)\,
\sum_{w_\2} \,F_{\si_\1\si_\2}(\rho_\1|\rho_2)_{w_\1}^{w_\2}\,
f_{w_\2}^{}(\rho_\2)\,.
\end{equation}
The conditions \rf{crossing} can be seen to be equivalent to the
following condition of invariance 
of the family of hermitian forms $H_\si$,
\begin{equation}\label{crossing2}
H_{\si_\1}^{}\big(\,{\mathsf M}_{\si_\1\si_\2}^{} f_{\si_\2}^{}\,,\,
{\mathsf M}_{\si_\1\si_\2}^{} g_{\si_\2}^{}\,\big)\,=\,
H_{\si_\2}^{}\big(\,f_{\si_\2}^{}\,,\,g_{\si_\2}^{}\,\big)\,.
\end{equation}
Combined with the factorization \rf{Hfactor} 
one gets nontrivial restrictions on the family
of hermitian forms $H^{\rho'\rho}$ on
${\mathfrak F}(\rho')\times
 {\mathfrak F}(\rho)$.

\subsection{Representation of the relations on spaces of conformal blocks}

\subsubsection{}

The operators $\SZ_p$, $\SB_p$, $\SF_{pq}$ and 
$\SS_p$ defined in the previous subsection are not independent. 
The faces $\varpi$ 
of the complex $\CM(\Sigma)$ correspond to round trips in ${\mathfrak T}_{g,n}$
starting and ending in the fundemantal 
domain $\CV_{\si}$ associated to some marking $\si$.
The realization of such round trip on the conformal blocks
can be realized by means of 
parallel transport w.r.t. the canonical connection on 
spaces of conformal blocks, in general.
The operator 
$\SU(\pi)$ which is associated to a round trip therefore has
to be proportional to the identity. It does not have to be equal to the
identity since the canonical connection is not flat but only projectively flat.
One thereby gets a relation ${\mathsf U}(\varpi)=\zeta_\varpi$ 
between the ``generators''
$\SZ_p$, $\SB_p$, $\SF_{pq}$ and 
$\SS_p$ for every round trip $\varpi$ that one can 
compose out of the elementary moves $(pq)$ 
$Z_p$, $B_p$, $F_{pq}$ or $S_p$.

\subsubsection{Insertions of the vacuum}



Taking into account the propagation of vacua
implies that the operators 
$\SZ_p$, $\SB_p$, $\SF_{pq}$
must simplify considerably if one of the representations
inserted is the vacuum. We must have, in particular, the 
relations
\begin{align}
& \SF\cdot\big[{\mathfrak H}_{\bar{r}_\3r_\3}^{0} \ot 
{\mathfrak H}_{r_\2r_\1}^{r_\3}\big]
=\SZ\cdot{\mathfrak H}_{r_\3r_\2}^{r_\1}\ot {\mathfrak H}_{\bar{r}_\1{r}_\1}^{0} 
\label{fustriv}\\
&\SF\cdot\big[{\mathfrak H}_{0 \bar{r}_\3}^{r_\3} \ot {\mathfrak H}_{r_\2r_\1}^{r_\3}\big]
={\mathfrak H}_{0r_\2}^{\bar{r}_\2}\ot {\mathfrak H}_{r_\2r_\1}^{r_\3} \\
& \SB\cdot{\mathfrak H}_{0r}^{\bar r}={\mathfrak H}_{r0}^{\bar r}\label{braidtriv}\\
& \SB\cdot{\mathfrak H}_{r\bar{r}}^{0}= \ST\cdot{\mathfrak H}_{\bar{r}r}^{0}\,.
\label{brtriv}\end{align} 
We are using the notation ${\mathfrak H}_{r_\2r_\1}^{r_\3}$ for 
$\CC(S_{\sst 0,3},\rho)$ if $\rho=(r_\3,r_\2,r_\1)$, with 
labelling of boundary components in correspondence to the decoration
introduced in Figure \ref{marking}.

Some of the 
relations in which these operators appear trivialize 
accordingly. 
By rescaling 
the operators $\SZ_p$, $\SB_p$, $\SF_{pq}$ and 
$\SS_p$, if neccessary, one can achieve that
$\zeta_\varpi=1$ for some other relations. 
With the help of these observations
it is easy to see that one may assume 
$\xi_\varpi=1$ for each of the relations \rf{zrel}-\rf{pentarel}
associated to Riemann surfaces of genus zero.

In the case of genus one let us observe that the 
relation \rf{twotorrel} trivializes if one of the
two external representations is the vacuum representation.
One may furthermore always redefine $\SS_p$ such that
$\xi_\varpi=1$ in the case of the relation which corresponds 
to \rf{onetor:a}, b). One is left with the relation  \rf{onetor:a}, a).
The arguments well-known from rational CFT \cite{MS,BK1} lead one to
the conclusion that the corresponding relation is
\begin{equation}
\SS_p^2\,=\,e^{-\pi i\frac{c}{2}}\,\SB_p'\,,
\end{equation}
where $c$ is the central charge of the Virasoro algebra introduced
in \rf{Vir}.

\section{Notion of a stable modular functor}
\setcounter{equation}{0}

We shall now formulate an abstract framework that 
we believe to be suitable for large classes of 
not necessarily rational CFT. This framework can be seen as a variant 
of the concept of a modular functor from \cite{S}. 
Combined with the gluing construction of the conformal
blocks it will be shown to yield a concrete realization 
of the point of view of Friedan and Shenker \cite{FS} 
who proposed to view the partition function of a CFT 
as a hermitian metric on a projective line bundle 
over the moduli space of Riemann surfaces, with 
expectation value of the stress-energy tensor being 
the canonical connection.

An important feature is that we will assume existence
of a scalar product on the spaces of conformal blocks.
The topology defined by the scalar product
gives us control on the possible infinite-dimensionality
of these spaces. 
Existence
of a scalar product
may seem to be an overly strong assumption,
but we will discuss in the following sections why 
we believe that the class of vertex algebras that 
is covered by our formalism  is rather large.

\subsection{Towers of representations of the modular groupoid}


\begin{defn}
A tower of representations of the modular groupoid assigns to 
a topological surface  $\Sigma$
the following objects:
\begin{itemize}
\item
the family of Hilbert spaces 
$\big[{\mathfrak H}_\si\big]_{\si\in \CM_\0(\Sigma)}$
of the form 
\begin{equation}
{\mathfrak H}_\si\,\equiv\,
\int\limits_{\BU^{\si_\1}}^{\oplus}  d\mu_\si(\rho)
\bigotimes_{p\in\si_\0}{\mathfrak F}(\rho_p) \,,
\end{equation}
where $d\mu_\si(\rho)$ is the product measure 
$d\mu_\si(\rho)=\prod_{e\in\si_\1}d\mu_{\rm\sst Pl}(r_e)$ if
$\rho:\si_\1\ni e\ra r_e\in\BU$.

\item For each pair $[\si_\2,\si_\1]$ of markings a unitary operator
$\SM_{\si_\2\si_1}:{\mathfrak H}_{\si_\1}\ra{\mathfrak H}_{\si_\2}$ such that 
\begin{equation}\label{MMM}
\SM_{\si_1\si_\3}\cdot\SM_{\si_3\si_\2}\cdot\SM_{\si_2\si_\1}\,=\,
\xi_{\si_\3\si_\2\si_\1}\cdot{\rm id}\,,
\qquad\SM_{\si_1\si_\2}\cdot\SM_{\si_2\si_\1}\,=\,1\,,
\end{equation}
where $\xi_{\si_\3\si_\2\si_\1}\in\BC$, $|\xi_{\si_\3\si_\2\si_\1}|=1$.
\end{itemize}
This assignment is such that the following 
requirements hold:
\begin{itemize}
\item[] {\bf Disjoint union:} 
Let $X=X'\sqcup X''$ be the disjoint union of $X'$ and
$X''$, and let  $\si_i=\si_i'\sqcup\si_i''$, $i=1,2$ be two markings
on $X$. Then
\begin{align}
& \CH_{\si}\,=\,\CH_{\si'}\ot \CH_{\si''}\,,\\
& \SM_{\si_\2\si_\1}\,=\,\SM_{\si_\2'\si_\1'}\ot \SM_{\si_\2''\si_\1''}\,.
\end{align}
\item[] {\bf Gluing:} Assume that 
$X'$ is obtained from $X$ by gluing the boundary
components $B_\2$ and $B_\1$. Let $\si$ be a marking on $X$ 
with edges $e_\2$ and $e_\1$ ending in $B_\2$ and $B_\1$,
respectively, and let $\si'$ be the marking on $X$ obtained from $\si$
by gluing $e_\2$ and $e_\1$.

There is a dense subset $\CT_\si$ of $\CH_\si$ such that for each 
$\Psi\in \CT_\si$ the following gluing projection
is well defined:\[ 
\SG_{e_\2e_\1}\Psi\,=\,
\Psi|_{s_{e_\2}=s_{e_\1}}^{}
\,,\]
where $\Psi|_{s_{e_\2}=s_{e_\1}}^{}$ is obtained from 
$\Psi$ by restricting it to the subset of $\CI_\si$ where $r_{e_\1}=r_{e_\2}$.

The projection $\SG_{e_\2e_\1}$ is then required
to be compatible with the representations of the modular
groupoids of $X$ and $X'$ in the following sense:
It is required that $\SM_{\si_\2\si_1}$
maps from $\CT_{\si_\1}$ to $\CT_{\si_\2}$ and that
\begin{equation}
\SG_{e_\2^{}e_\1^{}}\SM_{\si_\2\si_1}\Psi\,=\,
\SM_{\si_\2'\si_1'}\SG_{e_\2^{}e_\1^{}}\Psi\,,
\end{equation}
holds for all pairs $(\si_\2',\si_\1')$ of markings on $X$ obtained from the 
corresponding markings $(\si_\2,\si_\1)$ on $X$ by gluing, 
and all $\Psi\in\CT_{\si_\1}$. 
\item[] {\bf Propagation of vacua:} Let $X'$ be obtained from $X$
by gluing a disc into the boundary component $B_\0$. Let
$\si$ be a marking of $X$, $e_\0$
be the edge of $\si$ that ends in $B_\0$ and $p_\0$ be the vertex 
from which $e_\0$ emanates. One gets a marking $\si'$ on $X'$ 
by deleting $e_\0$ and $p_\0$ and gluing the 
other two edges that emanate from $p_\0$ into a single edge of $\si'$.

There then exist  dense subsets $\CT_{\si}$ and
 $\CT_{\si'}$ of $\CH_{\si}$ and $\CH_{\si'}$, respectively,  
together with projection mappings $P_{\si,B_\0}:\CT_{\si}\mapsto
\CT_{\si'}$ such that
\begin{equation}\label{propvac}
\SP_{\si_\2,B_\0^{}}\SM_{\si_\2\si_1}\Psi\,=\,
\SM_{\si_\2'\si_1'}\SP_{\si_\1,B_\0^{}}\Psi\,,
\end{equation}
holds for all pairs $(\si_\2,\si_\1)$ of markings on $X$ and the
corresponding markings $(\si_\2',\si_\1')$ on $X'$ defined above.
\end{itemize}
\end{defn}
The requirements concerning disjoint union and gluing imply 
that the representations of the modular 
groupoids are constructed out of the representatives of the 
elementary moves $B_p$, $Z_p$, $F_{pq}$ and $S_p$.
A tower of representations of the modular groupoid is therefore
characterized
by the following  data: 
\begin{itemize}
\item The measure set $\BU$ (labels of unitary representations of $V$), 
equipped with a measure $d\mu_{\rm\sst Pl}$.

\item The Hilbert spaces ${\mathfrak H}_\si^{{}^{(\3)}}(\rho)$ 
associated to the markings $\si$ on the
three punctured sphere $\Sigma_{0,3}$ with 
assignment $\rho:k\mapsto r_k$, $k\in\{1,2,3\}$. 
\item The operators $\SZ$, $\SB$, $\SF$ and $\SS$ mapping 
${\mathfrak H}_{\si_\1}(\rho)$ to ${\mathfrak H}_{\si_\2}(\rho)$ 
with respective 
markings $\si_\1$ and $\si_\2$ being chosen as depicted in 
figures \ref{zmove}-\ref{smove}.
\end{itemize}
Let us furthermore remark that the constraints imposed on these data 
by the propagation of vacua requirement are related to 
\rf{fustriv}-\rf{brtriv}. The
precise relationship can be subtle if the vacuum representation
is not contained in support of $d\mu_{\rm\sst PL}$ as it may
happen for nonrational CFT (see Section \ref{Liourev} for an example). 
The definition of the projection mappings $P_{\si,B_\0}$ then 
involves analytic continuation w.r.t. the conformal dimensions of the
representations, and the compatibility condition \rf{propvac} 
requires that the dependence of $\SZ$, $\SB$, $\SF$ and $\SS$
on the labels of external representations has a sufficiently large
domain to analyticity. 

\subsection{Unitary modular functors}

Given a tower of representations of the
modular groupoids 
there is a canonical way to construct a
corresponding modular functor, as we shall now explain. 
The main issue is to eliminate the apparent
dependence on the choice of the marking $\si$.

Each of the spaces ${\mathfrak H}_\si$ becomes a representation
of the mapping class group ${\rm MCG}(\Sigma)$ 
by choosing for each $m\in{\rm MCG}(\Sigma)$ a sequence 
$\pi_m$ of elementary moves that connects $\si$ to $m(\si)$. 
Taking advantage of the fact that the isomorphism
${\mathfrak H}_\si\simeq {\mathfrak H}_{m(\si)}$ is canonical
one gets an operator $\SM(m)$ on ${\mathfrak H}_{\si}$.

It is easily seen that for each pair $[\si_\2,\si_\1]$ 
there exist  numbers
$\zeta_{\si_\2\si_\1}^{}$ which satisfy 
\begin{equation}\label{cocycle}
\zeta_{\si_1\si_\3}\cdot\zeta_{\si_3\si_\2}\cdot\zeta_{\si_2\si_\1}\,=\,
\xi_{\si_\3\si_\2\si_\1}\cdot{\rm id}\,.
\end{equation}
Indeed, given a fixed reference marking $\si_\0$ one may take e.g.
$\zeta_{\si_\2\si_\1}^{}\equiv \xi_{\si_\2\si_\0\si_\1}^{-1}$. 
This means that one can use the numbers $\zeta_{\si_\2\si_\1}^{}$
to  define a {\it projective} 
holomorphic
line bundle $\CL_V$ 
over ${\mathfrak T}_{g,n}$. To this aim,
use the $\CV_{\si}$ as local coordinate patches, with 
transition functions $\zeta_{\si_\2\si_\1}^{}$. Projectiveness
follows from the nontriviality of the phase
$\xi_{\si_\3\si_\2\si_\1}$ associated to triples of markings.

\begin{defn}\label{Global} Let ${\mathfrak H}(\Sigma)$ be the Hilbert space
whose elements $\Phi$ 
are collections of vectors $\Psi_\si\in{\mathfrak H}_\si$ 
such that
\begin{equation} \label{equiv}
\Psi_{\si_\2}\,=\,  \zeta_{\si_\2\si_\1}^{-1}\,
\SM_{\si_\2\si_\1}^{}\Psi_{\si_\1}^{}\,,
\end{equation}
holds for all pairs of markings $\si_\1$, $\si_\2$ 
and a given collection of 
complex numbers $\zeta_{\si_\2\si_\1}^{}$ of modulus one 
which satisfy \rf{cocycle}.

For a given collection of
numbers $\eta_\si\in\BC$, $\si\in\CM_{\0}(\Sigma)$, $|\eta_\si|=1$, let us call the operation
$\Psi_\si\ra \eta_\si\Psi_\si$ for all $\Psi_\si\in{\mathfrak H}_\si$
a gauge transformation. We will identify the Hilbert spaces
${\mathfrak H}(\Sigma)$ related by gauge transformations.

Let $\rho(\Sigma)$ be 
the family of mapping class group representations 
$(\rho_\si(R))_{\si\in \CM_\0(\Sigma)}$ on the
spaces ${\mathfrak H}_\si(R)$ modulo the equivalence
relation $\sim$ that is induced by the identifications \rf{equiv}.

The assignment $\Sigma\ra ({\mathfrak H}(\Sigma),\SM(\Sigma))$ 
will be called a stable unitary projective functor.
\end{defn}

\subsection{Similarity of modular functors}

For rational CFT there exist deep results on the equivalence
of modular functors from conformal field theory to similar objects
coming from quantum group theory \cite{Fi}. In order to formulate analogous
statements about nonrational CFT we will propose the following
natural notion of similarity of modular functors.

\begin{defn}\label{simil}
We will call two modular functors $\CF$ and $\CF'$ with data
\begin{align*}
&\big[\,\BU\phantom{'}\,,\,d\mu_{\rm\sst Pl}\,,\,
{\mathfrak H}_\si^{\sst (3)}(\rho)\,,\,
\SZ\phantom{'}\,,\,\SB\phantom{'}\,,\,\SF\phantom{'}\,,\,
\SS\phantom{'}\,\big]\\
&\big[\,\BU'\,,\,d\nu_{\rm\sst Pl}\,,\,{\mathfrak K}_\si^{\sst (3)}(\rho')\,,\,
\SZ'\,,\,\SB'\,,\,\SF'\,,\,\SS'\,\big]
\end{align*}
similar iff the following conditions are satisfied:
\begin{itemize}
\item There exists 
a bijection between $\BU$ and $\BU'$. The measures $d\mu_{\rm\sst Pl}$ 
and $d\nu_{\rm\sst Pl}$ are 
equivalent, 
i.e. there exists a  positive function $m(r)$
on $\BU$ such that 
\[
d\mu_{\rm\sst Pl}(r)\,=\,m(r)d\nu_{\rm\sst Pl}\,.
\]
\item There exist families of invertible 
operators $\SE^{\sst 0,3}(\rho): {\mathfrak H}^{\sst (3)}(\rho)\ra
{\mathfrak K}^{\sst (3)}(\rho')$, the dependence on each representation
label $r_k\in\BU$, $k=1,2,3$
measurable w.r.t. $d\mu_{\rm\sst Pl}(s)$ such that the operators 
$\SE_\si:{\mathfrak H}_\si\ra {\mathfrak K}_\si$ defined as
\begin{equation}
\SE_\si\,\equiv\,\int\limits_{\BU^{\si_\1}}^{\oplus}d\nu_{\sst \si}(\rho)
\;\prod_{p\in\si_\0}
\SE^{\sst 0,3}_{}(\rho_p)
\end{equation}
are invertible.
\item
The resulting operators $\SE_\si:{\mathfrak H}_\si\ra
{\mathfrak K}_\si$ satisfy
\[
\SM_{\si_\2\si_\1}'\,=\,
\SE_{\si_\2}^{}\!\!\cdot \SM_{\si_\2\si_\1}^{}\!\!\cdot \SE_{\si_\1}^{-1}\,.
\]
\end{itemize}
\end{defn}

\subsection{Friedan-Shenker modular geometry}

Let us temporarily restrict attention to surfaces $X$ with one marked
point at position $z\in X$, decorated with the vacuum representation $V$. 
We will assume that the values of the conformal
blocks $G_{\si\tau}(\de|v)$ 
at a given vector $v\in V$ 
may be considered as a family $(G_{\si\tau}(v))_{\tau\in\CV_{\si}}$ 
of elements 
of the Hilbert space ${\mathfrak H}_\si$. 

Out of  $(G_{\si\tau}(v))_{\tau\in\CV_{\si}}$ one may
then define a collection of vectors 
$\{\,\Psi_{\si;\si'}(v|\tau)\,;\,\si'\in\CM_\0(\Sigma)\,\}$, where 
$\Psi_{\si;\si'}(v|\tau)\in
{\mathfrak H}_{\si'}$ for all $\si'\in\CM_\0(\Sigma)$ 
such that the conditions 
\begin{equation}\label{Psiglobal}
\Psi_{\si;\si_\2}(v|\tau)\,=\,\zeta_{\si_\2\si_\1}^{-1}\,
\SM_{\si_\2\si_\1}^{}\Psi_{\si;\si_\1}^{}(v|\tau)\,\quad{\rm and}\quad
\Psi_{\si;\si}(v|\tau)\,=\,G_{\si\tau}(v)\,\in\,{\mathfrak H}_\si
\end{equation}
are satisfied. Indeed, consistency of the definition of 
 $\Psi_{\si;\si'}(v|\tau)$ implied by \rf{Psiglobal} follows from \rf{MMM} and
\rf{cocycle}.
Let  $(\Phi_{\si\tau}(v))_{\tau\in\CV_{\si}}$ 
be the holomorphic
family of vectors
in ${\mathfrak H}(\Sigma)$ which is associated 
by Definition \ref{Global} to the collection 
$\{\,\Psi_{\si;\si'}(v|\tau)\,;\,\si'\in\CM_0(\Sigma)\,\}$.

Given two markings $\si_\2$, $\si_\1$ such that 
$\CV_{\si_\2}\cap \CV_{\si_\1}\neq\emptyset$ 
it is easy to see that the families
of vectors  $(\Phi_{\si_\2\tau_\2}(v))_{\tau_\2\in\CV_{\si_\2}}$ 
and $(\Phi_{\si_\1\tau_\1}(v))_{\tau_\1\in\CV_{\si_\1}}$ 
are related as
\begin{equation} \label{transpsi}
\Psi_{\si_\2\tau}^{}(v)\,=\,\zeta_{\si_\2\si_\1}^{}
\Psi_{\si_\1\tau}^{}(v)\,
\end{equation}  if $\tau_\2$ and $\tau_\1$ 
parametrize the same point in 
$\CV_{\si_\2}\cap \CV_{\si_\1}$. 
Indeed,
we had defined $\SM_{\si_\2\si_\1}$ in \rf{genfusion},\rf{Fdef} 
such that $G_{\si_\1\tau_\1}(v)
{=}{\mathsf M}_{\si_\1\si_\2}
G_{\si_\2\tau_\2}(v)$.
This implies  \[
\Psi_{\si_\1;\si_\1}(v|\tau_\1)=
G_{\si_\1\tau_\1}(v)
{=}{\mathsf M}_{\si_\1\si_\2}
G_{\si_\2\tau_\2}(v)={\mathsf M}_{\si_\1\si_\2}
\Psi_{\si_\2;\si_\2}(v|\tau_\2)\overset{(\ref{equiv})}{=}\zeta_{\si_\1\si_\2}
\Psi_{\si_\2;\si_\1}(v|\tau_\2)\,.
\]
This
means that for each $\si$ one may regard the family 
$(\Psi_{\si\tau}(v))_{\tau\in\CV_{\si}}$ as
a local holomorphic section of the  projective line bundle  $\CL_V$ over  
${\mathfrak T}_{g,n}$.

The invariance conditions \rf{crossing2} imply that
the family of hermitian forms  $H_\si$ defines a hermitian form
$H$ on ${\mathfrak H}(\Sigma)$. 
Objects of particular interest for the case at hand are the 
partition function $Z_g(X)$,
\begin{equation}
Z_g(X)\,\equiv\,H\big(\,\Psi_{\si\tau}(v_\0)\,,
\,\Psi_{\si\tau}(v_\0)\,\big)
\end{equation}
and the expectation values $\bra\!\bra\, Y(A,z)\,\ket\!\ket$ of 
local fields $ Y(A,z)$ 
from the vertex algebra 
$V$,
\begin{equation}
\bra\!\bra\, Y(A,z)\,\ket\!\ket\,\equiv\,\frac{H\big(\,\Psi_{\si\tau}(A)\,,
\,\Psi_{\si\tau}(A)\,\big)}{H\big(\,\Psi_{\si\tau}(v_\0)\,,
\,\Psi_{\si\tau}(v_\0)\,\big)}\,.
\end{equation}

Following \cite{FS} we will regard the partition function 
$Z_g(X)$ as a hermitian metric $\CH$ on the projective 
line bundle $\CL_V^{}$.  
It follows easily from \rf{Viract} that 
\begin{equation}\label{FSconn}
\de_\vt^{}\log Z_g(X)\,=\,
\big\langle\!\big\langle\, Y(T[\eta_\vt]v_\0,z)\,\big\rangle\!\big\rangle\,,
\end{equation}
where $T[\eta]=\sum_{n\in\BZ}\eta_nL_n$, 
$\de_{\vartheta}$ is the derivative corresponding 
a tangent vector $\vartheta\in T{\mathfrak M}_{g,0}$
and $\eta_\vartheta^{}$ 
is any element of $\BC(\!(t)\!)\pa_t$
which represents $\vartheta$ via \rf{VirUni}. 
Equation \rf{FSconn} can be seen as a more precise formulation of
the claim from \cite{FS} that the expectation value of the 
stress-energy tensor 
coincides with the connection on the projective 
line bundle $\CL_V^{}$ which is canonically associated with
the metric $\CH$.
We have thereby reconstructed the main ingredients of the formulation
proposed by
Friedan and Shenker \cite{FS} within the framework provided 
by the gluing construction.

\section{Example of a 
nonrational modular functor} \label{Liourev}
\setcounter{equation}{0}

There is considerable evidence for the claim that the most basic
example of a vertex algebra, the Virasoro algebra, yields a 
realization of the framework above.
The results of \cite{TL} are essentially equivalent to the construction
of the corresponding modular functor in genus 0. In the following
section we shall review the main characteristics of this
modular functor.

\subsection{Unitary positive energy representations of the 
Virasoro algebra}

The unitary highest weight representations $R_{\De}$ of the Virasoro algebra
are labelled by the eigenvalue $\De$ of the Virasoro generator
$L_0$ on the highest weight vector. It will be convenient to
parametrize $\De$ as follows
\begin{equation}\label{alDe}
\De_\al\,=\,\al(2\de-\al), \qquad {\rm where}\qquad c=1+24\de^2\,.
\end{equation}
The representations $R_\al\equiv R_{\De_\al}$ 
are unitary iff $\De\in[0,\infty)$. In terms
of the parametrization \rf{alDe} one may cover this range by assuming that
\begin{equation}\label{BUdef}
\al\,\in\, {\mathbb U}\,\equiv\,[0,\de]\,\cup\, \big( \de+i\BR^+\big)\,.
\end{equation}
The representation $R_\al$ for $\al=0$
corresponds to the vacuum representation $V$. The set parametrizes the unitary
dual of the Virasoro algebra. In order to indicate 
an important analogy with the 
representation theory of noncompact Lie groups we shall call the
family of representations $R_\al$ with $\al\in\de+i\BR^+$
the principal series of representations, 
which constitute the tempered dual $\BT$ of the 
Virasoro algebra.
Pursusing these analogies it seems natural to 
call the  family of representations
$R_\al$ with $\al\in [0,\de]$  
the complementary series. 

\subsubsection{Free field representation}

The Fock space $\CF$ is defined to be the 
representation of the commutation relations
\begin{equation}\label{ccr0}
[\sa_n,\sa_m]=\frac{n}{2}\de_{n+m}\,, 
\end{equation}
which is generated from the vector $\Omega\in\CF$ characterised by 
$a_n\Omega=0$ for $n>0$. There is a unique scalar product
$(\,.\,,\,.\,)_\CF^{}$ on $\CF$ such that $\sa_n^{\dagger}=\sa_{-n}^{}$ 
and $(\,\Omega\,,\,\Omega\,)_\CF^{}=1$.

Within $\CF$ we may define a one-parameter family
of representations $\CF_p$ of the Virasoro
algebra by means of the formulae
\begin{equation}
\begin{aligned}
& L_n(p)\,=\,2(p+in\de)a_n+\sum_{k\neq 0,n}a_ka_{n-k}\,,\qquad
n\neq 0\,,\\
& L_0(p)\,=\,p^2+\de^2+2\sum_{k>0}a_{-k}a_{k}\,.
\end{aligned}
\end{equation}
The representation $\CF_p$  is 
unitary w.r.t. the scalar product  $(\,.\,,\,.\,)_\CF^{}$
if $p\in\BR$. It is furthermore 
known \cite{Fr} to be irreducible and 
therefore isomorphic to  $R_\al$ if 
$p=-i(\al-\de)$ for all $\al\in\BU$.

\subsection{Construction of Virasoro conformal blocks in genus zero}

In the case of the Virasoro algebra there exists a unique 
conformal block $G$ associated to the three punctured
sphere which satisfies
\begin{equation}\label{F3norm}
G(\,\rho\,|\,{v}_{\al_\3}\ot {v}_{\al_\2}\ot
{v}_{\al_\1})\,=\,1\,,
\end{equation}
$v_{\al_k}$, $k\in\{1,2,3\}$ being the highest weight vectors of the
representations $R_{\al_k}$, respectively.
The corresponding family of operators 
$\SY_{\al_\3\al_\1}^{\al_\2}({\mathfrak v}_\2|z):
R_{\al_\1}\ra R_{\al_\3}$ is uniquely characterized 
by its member corresponding to ${\mathfrak v}_\2=v_{\al_2}$,
which will be denoted $\SY_{\al_\3\al_\1}^{\al_\2}(z)$.

\subsubsection{Free field construction of chiral vertex operators}

Let us introduce the (left-moving) 
chiral free field $\vf(x)\;=\;\sq+\spp x + 
\vf^{}_<(x)+\vf^{}_>(x),$ by means of the expansions
\begin{equation}
\vf_{<}(x)\;=\; 
i\sum_{n< 0}
\frac{1}{n}\, \sa_n \, e^{-inx},
\qquad
\vf_{>}(x)\;=\;
i\sum_{n> 0}
\frac{1}{n}\,\sa_n \,e^{-inx},
\end{equation}
The operators $\sq$ and $\spp$ are postulated to have the following 
commutation and hermiticity relations
\begin{equation}\label{ccr}
[\sq,\spp]=\frac{i}{2},\qquad
\sq^{\dagger}= \sq,\qquad
\spp^{\dagger}=\spp,
\end{equation}
which are naturally realized in the  
Hilbert-space
\begin{equation}
{\mathfrak H}^{\rm\sst F}\;\equiv \;L^2(\BR)\ot\CF\,.
\end{equation}
Diagonalizing the operator $\spp$ corresponds to decomposing 
${\mathfrak H}^{\rm\sst F}$ as direct integral of irreducible unitary
representations of the
Virasoro algebra,
\begin{equation}\label{irrdecomp}
\CM\;\simeq\;\int_{\BT}^{\oplus}d\al \;R_{\al}\,.
\end{equation}

The basic building blocks of all constructions will be the following 
objects:\\[1ex]
\noindent{\sc Normal ordered exponentials : }
\begin{equation} 
\SE^{\al}(x)\;
\equiv\;
\SE^{\al}_{<}(x)\SE^{\al}_{>}(x),
\qquad 
\begin{aligned}
{} & 
\SE^{\al}_<(x)\;=\;
e^{\al \sq}\;
e^{2\al\vf^{{+}}_{<}(x)}\;
e^{\al x \spp}\\
{} & 
\SE^{\al}_>(x)\;=\;
e^{\al x \spp}\;
e^{2\al\vf^{{+}}_{>}(x)}\;
e^{\al\sq} 
\end{aligned}
\end{equation}
{\sc Screening charges:}
\begin{equation}
\SQ(x)\;\equiv\;e^{-\pi b\spp}
\int_0^{2\pi}dx'\;\SE^b(x+x')\;e^{-\pi b\spp}\;.
\end{equation}

The following property is of considerable importance:\\[1ex]
{\sc Positivity: } The screening charges are densely defined
positive operators, i.e.
\[
\big(\, \psi\,,\,Q(\si)\psi\,\big)_{\CM}\;\geq\;0\;
\]
holds for $\psi$ taken from a dense subset of ${\mathfrak H}$.

Out of the building blocks introduced in the previous subsection 
we may now construct an important class of chiral fields,
\begin{equation}
\sh_s^{\al}(\si)\;=\;
\SE^{\al}(\si)\,
\big(\SQ(\si)\big)^s
\;,\end{equation}
Positivity of $\SQ$ allows us to consider these
objects for {\it complex} values of $s$ and $\al$.

One of the most basic properties of the $\sh_s^{\al}(w)$ are the
simple commutation relations with functions of the operator $\spp$,
\begin{equation}\label{shift}
\sh_s^{\al}(w)f(\spp)\;=\;f\big(\spp-i(\al+bs)\big)
\sh_s^{\al}(w)\,.
\end{equation}
By projecting onto eigenspaces of $\spp$ one may therefore
define a family of operators $\sh_{\al_\3\al_1}^{\al_2}(w):R_{\al_\1}\ra
R_{\al_\3}$. Specifically, for each ${\mathfrak w}\in\CF$ and each 
$\al\in\de+i\BR$
let us define a distribution  ${\mathfrak w}_\al$
on 
dense subspaces of $\CM$ by the relation
$(\,{\mathfrak w}_{\al}\,,\,{\mathfrak v}\,
)^{}_{\CM}=
(\,{\mathfrak w}\,,\,{\mathfrak v}_{\al}\,
)^{}_{\CF}$ if ${\mathfrak v}$ is represented via
\rf{irrdecomp} by the family of vectors
${\mathfrak v}_{\al}$, $v_\al\in R_\al$.
This implies that the matrix elements of the operators
 $\sh_{\al_\3\al_1}^{\al_2}(w)$ are determined by the relation
\begin{equation}
\big(\,{\mathfrak w}\,,\,
\sh_{\al_\3\al_1}^{\al_2}(w)\,{\mathfrak u}\,\big)_{\CF}^{}
\,=\,
\big(\,{\mathfrak w}_{\al_\3}\,,\,\sh_s^{\al_\2}(w)\,\hat{\mathfrak u}\,
\big)^{}_{\CM}\,,
\end{equation}
where $bs=\al_\3-\al_\1-\al_\2$ and $\hat{\mathfrak u}$ is 
any vector in $\CM$ represented by the family of vectors
$\hat{\mathfrak u}_{\al}$ such that $\hat{\mathfrak u}_{\al_\1}={\mathfrak u}$.

The uniqueness of the 
conformal block $G^{\sst (3)}$  implies that the
operator ${\mathsf Y}_{\al_\3\al_\1}^{\al_\2}({\mathfrak v}_\2|\,z\,)$ 
must be proportional
to $\sh_{\al_\3\al_\1}^{\al_\2}({\mathfrak v}_\2\,|\,z)$ via
\begin{equation}
{\mathsf Y}_{\al_\3\al_\1}^{\al_\2}({\mathfrak v}_\2|\,z\,)
\,=\,N_{\al_\3\al_\1}^{\al_\2}\,
\sh_{\al_\3\al_\1}^{\al_\2}({\mathfrak v}_\2\,|\,z)\,.
\end{equation}
The explicit formula for the
normalizing factor  $N_{\al_\3\al_\1}^{\al_\2}$ was found in \cite{TL}.

\subsection{Factorization property}

The results of \cite{TL} show that the conformal blocks
in genus zero satisfy the factorization property with linear
relations \rf{genfusion} composed from the elementary transformations
$F_{pq}$, ${B}_p$ and ${Z}_p$ whose representatives can
be calculated explicitly.

\begin{itemize}

\item[{\bf F}:] Let $X$ be a four-punctured sphere and let $\si_s$, $\si_u$
be the two markings depicted in Figure \ref{fmove}. 
We will denote the respective assignments of representation
labels to the edges of $\si_s$ and $\si_u$ by
$\rho_s(\al_s)$ and $\rho_u(\al_u)$, respectively, leaving implicit
the assignment of labels $\al_\1,\al_\2,\al_\3,\al_\4$ to the
external edges with numbering being indicated in Figure \ref{fmove}. 
The operator ${\mathbf F}_{pq}$ may then be represented as the
integral operator
\begin{equation}\label{LiouF}
G_{\si_s\tau_\2}(\rho_s(\al_s))
\,=\,\int d\mu_{\rm \sst Pl}(\al_u)\;{\rm F}_{\al_s\al_u}\big[{}_{\al_\4}^{\al_\3}{}_{\al_\1}^{\al_\2}\big]\,G_{\si_u\tau_\1}(\rho_u(\al_u))\,.
\end{equation}
The explicit expression for the kernel
$\displaystyle
{\rm F}_{\al_s\al_u} \big[ {}_{\al_\4}^{\al_\3} {}_{\al_\1}^{\al_\2} \big]$
can be found in \cite{TR,TL}.
More illuminating is probably the observation that the 
kernel ${\rm F}_{\al_s\al_u}$ is closely related to the 
$6j$ symbols of the modular double \cite{Fa,BT} 
of $\CU_q({\mathfrak s\mathfrak l}(2,\BR))$,
\begin{equation}
{\rm F}_{\al_s\al_u} \big[ {}_{\al_\4}^{\al_\3} {}_{\al_\1}^{\al_\2} \big]
\,=\,
\frac{ {\nu_{\al_\3\al_s}^{\al_\4}}
{\nu_{\al_\2\al_\1}^{\al_s}} }{ {\nu_{\al_u\al_\1}^{\al_\4}}
{\nu_{\al_\3\al_\2}^{\al_u}} }\,
\big\{ {}^{\al_1}_{\al_3} {}^{\al_2}_{\al_4}\big|{}_{\al_u}^{\al_s}\big\}\,.
\end{equation}
The explicit formula for the normalizing factors $\nu_{\al_\2\al_\1}^{\al_s}$
can be found in \cite{TR}.

\item[{\bf B}:] Let $X$ be a three-punctured sphere and let $\si_\2$, 
$\si_\1$ be the two markings depicted on the left and right halves of 
Figure \ref{bmove}, respectively. Let $\rho$ be the assignment 
$\rho:k\ra \al_k$, $k=1,2,3$  of representation labels to edges
as numbered  in Figure \ref{bmove}. We then have
\begin{equation}
G_{\si_\2}(\rho)\,=\,{\rm B}_{\al_3\al_2\al_1}G_{\si_\1}(\rho)\,,\qquad 
{\rm B}_{\al_3\al_2\al_1}\,\equiv\,
e^{\pi i(\al_\3(Q-\al_\3)-\al_\1(Q-\al_\1)-\al_\2(Q-\al_\2))}\,.
\end{equation}
\item[{\bf Z}:] Z is represented by the identity operator.
\end{itemize}
An important part of the statements above may be reformulated 
as the claim that the modular functor $\CF_{\rm\sst Teich}$ 
is similar in the sense of Definition \ref{simil} to a
modular functor $\CF_{\rm\sst Qgrp}$ that is constructed in close analogy
to the construction of Reshetikhin-Turaev from the representations 
of the modular double of $\CU_q({\mathfrak s\mathfrak l}(2,\BR))$
introduced in \cite{PT1,Fa} and studied in more detail 
in \cite{PT1,BT}.

\subsection{The Hilbert space structure}

As explained above, we need a pair
$\big[
\,
{\mathfrak H}_\si^{\sst (3)}(\rho)\,,\,d\mu_{\rm\sst Pl}\,\big]$ 
of objects in order to characterize the Hilbert space structures on the 
spaces of conformal blocks.

{\bf Hilbert space ${\mathfrak H}_\si^{\sst (3)}(\rho)$:}
It is well-known that the space of conformal blocks on the three punctured
sphere
is at most one-dimensional. More precisely we have:
\begin{equation}
{\mathfrak H}_\si^{\sst (3)}(\rho)\,\simeq\,\BC, \qquad \rho:k\mapsto 
\al_k, \;\;k\in\{1,2,3\}\,,
\end{equation}
if $\al_i\neq 0$ for $i=1,2,3$. If $\al_i=0$ for some $i\in\{1,2,3\}$ 
and if $k,l\in\{1,2,3\}$ are not equal to $i$ we have 
\begin{equation}
{\mathfrak H}_\si^{\sst (3)}(\rho)\,
\simeq\,\left\{\begin{aligned} & \BC\;\;{\rm if}\;
\al_k=\al_l \;\;{\rm or}\;\;\al_k=\2\de-\al_l\,,\\
&\emptyset\;\; {\rm otherwise.}
\end{aligned}\right.
\end{equation}
As a standard basis for ${\mathfrak H}_\si^{\sst (3)}(\rho)$ we shall use the 
conformal block $G^{\sst (3)}(\rho)$ that is uniquely defined by
the normalization \rf{F3norm}.
The Hilbert space structure on the one-dimensional
space ${\mathfrak H}_si^{\sst (3)}(\rho)$ is then described by
the numbers
\begin{equation}
D(\al_3,\al_2,\al_1)\,\equiv\,\lVert\, G^{\sst (3)}(\rho)\,\rVert^2\,,
\end{equation}
that are given explicitly by the formula
\begin{equation}
D(\al_3,\al_2,\al_1)\,=\,\left|\frac{\Ga_b(\al_{123}-Q)}{\Ga_b(Q)}
\prod_{k=1}^{3}
\frac{\Ga_b(\al_{123}-2\al_k)}{\Ga_b(2\al_k)}\right|^2\,,
\end{equation}
where $\Ga_b(x)\equiv
\Ga_2(x|b,b^{-1})$ with $\Ga_2(x|\omega_1,\omega_2)$ being 
the Barnes Double Gamma function, and
we have used the abbreviation $\al_{123}=\al_1+\al_2+\al_3$.

The measure $d\mu_{\rm \sst Pl}$ on $\BU$ will then be equal to
\begin{equation}
d\mu_{\rm \sst Pl}(\al)\,=\,d\al 
 \sin (2b(\al-\de))\,\sin (2b^{-1}(\de-\al))\quad{\rm on}\;\;\de+i\BR^+\,,
\end{equation} 
with $d\al$ being the standard Lebesque measure on 
$\de+i\BR^+$.

\subsection{Extension to higher genus}

\begin{claim}
There exists a unique extension of the $g=0$ modular functor 
$\CF_{\rm\sst Vir}$ to $g>0$ that is compatible with the propagation
of vacua. 

\end{claim}

The proof of this claim has not appeared in the literature
yet. Let us therefore briefly sketch the path along which the
author has arrived at the claim above. 

The main observation to be made is that there exists a 
unitary modular functor $\CF_{\rm\sst Teich}$ whose 
restriction to $g=0$ is similar to $\CF_{\rm\sst Vir}$
in the sense of Definition \ref{simil}. 
$\CF_{\rm\sst Teich}$ was constructed
in \cite{TT}.\footnote{The key step in the verification of our claim
above is to notice that the restriction 
of $\CF_{\rm\sst Teich}$ to surfaces with $g=0$
is similar to the modular functor
$\CF_{\rm\sst Qgrp}^{\sst g=0}$ coming from the 
harmonic analysis on the modular double of 
$\CU_q({\mathfrak s\mathfrak l}(2,\BR))$ that was mentioned above.
This establishes the  existence of an extension of 
$\CF_{\rm\sst Vir}$ to $g>0$.}

Uniqueness is in fact quite easily seen by noting that 
arguments well-known from rational conformal field theory \cite{MS2}
carry over to the case at hand and allow us to derive 
an explicit formula for the coefficients $S_{\al\be}(\ga)$
in terms of $\displaystyle
{\rm F}_{\al_s\al_u} \big[ {}_{\al_\4}^{\al_\3} {}_{\al_\1}^{\al_\2} \big]$
and ${\rm B}_{\al_3\al_2\al_1}$, namely
\begin{equation}\label{Sformula}\begin{aligned}
F_{0\be} & 
\big[ {}_{\al_\1}^{\al_\1} {}_{\al_\1}^{\al_\1} \big]
{S_{\al_\1\al_\2}(\be)}\,=\,\\
&=\,{S_{0\al_\2}(0)}
\int d\mu_{\rm\sst PL}^{}
(\ga) \;e^{\pi i(2\De_{\al_\2}+2\De_{\al_\1}-2\De_\ga-\De_\be)}\;
{{\rm F}_{0\ga} \big[ {}_{\al_\2}^{\al_\2} {}_{\al_\1}^{\al_\1} \big]}
{\rm F}_{\ga\be} \big[ {}_{\al_\2}^{\al_\1} {}_{\al_\2}^{\al_\1} \big]
\,.
\end{aligned}\end{equation}

\subsection{Remarks}


It is often natural to first focus attention on the
subspace of ``tempered'' conformal blocks which are obtained
from the gluing construction by using three point conformal
blocks associated to representations from the tempered 
dual $\BT$ only. The formulation of the theory as a
modular functor applies straightforwardly to this case.


However, in the  case of the Virasoro algebra  we may
observe rather nice analytic properties
of the conformal blocks when considered 
as functions of the representation labels $\al_k$ \cite{TR}.
The dependence w.r.t. the external representations
is {\it entire analytic}, while the dependence w.r.t. the internal 
representations is {\it meromorphic}.
The poles are given by the zeros of the Kac determinant.

The factorization property of the analytically continued conformal
blocks follows from the corresponding property of the
tempered conformal blocks. Analytic continuation w.r.t.
the representation labels therefore allows us to 
generate a large class of conformal blocks with 
reasonable behavior at the boundaries of the Teichm\"uller
spaces from the tempered conformal blocks. We will call this
class of conformal blocks the factorizable conformal blocks.
It is not clear to the author how this class compares
to the set of {\it all} solutions to the conformal
Ward identities.

\section{Existence of a canonical scalar product?}\label{canhermform}
\setcounter{equation}{0}

We propose that for each vertex
algebra $V$ whose conformal blocks have the 
factorization property there always exists a distinguished 
choice for $H_\si$,  canonically
associated to $V$, which  is ``diagonal'', i.e. such 
that $H_\si$ is of the form
\begin{equation}\label{scprod}
H_\si\,=\,\int\limits_{\BU^{\si_\1}}\prod_{e\in\si_\1}
d\mu_{\rm\sst Pl}(r_e)\;\bigotimes_{p\in\si_\0}
H_p^{\bar{\rho}_p\rho_p}\,,
\end{equation}
where $\bar{\rho}_p$ is the decoration of $S_p$ obtained from $\rho_p$ by
replacing each representation by its dual,
$\si_\1$ is the set of edges and
$\si_\0$ is the set of vertices of the graph $\si$. In rational
CFT this case is often referred to as the CFT corresponding to
the ``diagonal modular invariant''. 
We propose the terminology ``V-minimal model'' for the corresponding 
CFT. 

Whenever the hermitian form
$H$ is positive definite we get
a {\it scalar product} on the space of conformal blocks. We will 
subsequently
argue that this is always the case if the representations
in question are unitary.

\subsection{Existence of a canonical hermitian form from the factorization 
property}\label{HermFfactor}

\begin{figure}[t] 
\epsfxsize12cm
\centerline{\epsfbox{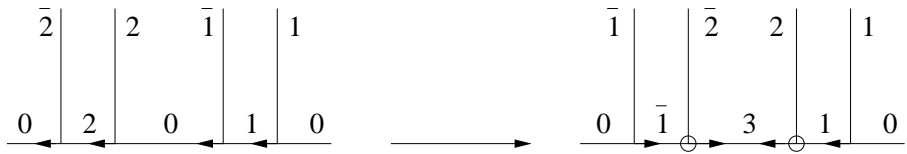}}
\caption{Simplified representation for the markings involved in the
relation \rf{spfus}.}
\label{fnullfig}\end{figure}

Note that ${\rm dim}{\mathfrak H}_{r_\2r_\1}^{r_\3}=1$ whenever one of the
representations $R_{r_i}$, $i=1,2,3$ coincides with the vacuum
representation, and the two other representations are $R$ 
and $\bar{R}$, with $\bar R$ being the dual of 
$R$. This implies that there is a unique (up to a constant)
conformal block associated to the diagram on the left of Figure \ref{fnullfig}
if the representation associated to the edges with 
label 0 is the vacuum representation and
if the representations associated to the edges with 
labels
$1,\bar{1},2,\bar{2}$ are chosen as $R_1,\bar{R}_\1,R_2,\bar{R}_2$,
respectively. This conformal block will be denoted as
$G^{\si_\1}_{\0\tau}\big[
\begin{smallmatrix} {r_\2} & {\bar{r}_\1}\\ {\bar{r}_\2} & r_\1
\end{smallmatrix}\big]$.

Let us, on the other hand, use the notation
$G_{r_\3\tau}^{\si_\2}\big[
\begin{smallmatrix} {\bar r_\2} & {{r}_\2} \\ {\bar{r}_\1} & r_\1
\end{smallmatrix}
\big]_{\imath\bar{\imath}}$ 
for the conformal blocks associated to the diagram on the
right of Figure \ref{fnullfig} in the case that the 
representations associated to the edges with 
labels
$1,2,\bar{1},\bar{2}$ are chosen as above. 
The indices $\imath$, $\bar\imath$ are associated to the 
vertices enclosed in little circles in a manner that 
should be obvious.

Bear in  mind that we are considering vertex algebras
whose conformal blocks have the factorization property. It follows
in particular that
the conformal blocks $G^{\si_\1}_{\0\tau}\big[
\begin{smallmatrix} {r_\2} & {\bar{r}_\1}\\ {\bar{r}_\2} & r_\1
\end{smallmatrix}\big]$ and $G_{r_\3\tau}^{\si_\2}\big[
\begin{smallmatrix} {\bar r_\2} & {{r}_\2}\\ {\bar{r}_\1} & r_\1
\end{smallmatrix}
\big]_{\bar\imath{\imath}}$
are related by an expansion of the form
\begin{equation}\label{spfus}
G^{\si_\1}_{\0\tau}\big[
\begin{smallmatrix} {r_\2} & {\bar{r}_\1}\\ {\bar{r}_\2} & r_\1
\end{smallmatrix}\big]
\,=\,
\int d\mu_{\1\2}(r_\3)\;\sum_{\imath,\bar{\imath}\in\BI_p}\;
D_{\bar\imath{\imath}}(r_\3|r_\2,r_\1)\,
G_{r_\3\tau}^{\si_\2}\big[
\begin{smallmatrix} {\bar r_\2} & {{r}_\2}\\ {\bar{r}_\1} & r_\1
\end{smallmatrix}
\big]_{\bar\imath{\imath}}\,.
\end{equation}

\begin{conj} \label{hermformconj}  $\quad$ \\
There exists a subset $\BT$ of  $\BU$ 
parametrizing ``tempered'' representations such that
for $r_1,r_2\in\BT$ the  measure $d\mu_{\1\2}$  is supported 
in $\BT$. In this case there exists a
factorization of the form
\begin{equation}\label{factmupl}
d\mu_{\1\2}(r_\3) \;
D_{\imath\bar{\imath}}(r_\3|r_\2,r_\1)\,=\,
d\mu_{\rm\sst Pl}(r_\3)\;
D_{\bar\imath{\imath}}(r_\3,r_\2,r_\1)\,,
\end{equation} 
with $d\mu_{\rm\sst Pl}$ being independent of $r_\2$, $r_\1$
such that 
the hermitian forms 
on spaces of conformal blocks constructed via \rf{scprod} from 
$d\mu_{\rm\sst Pl}(s)$ and $D_{\bar\imath{\imath}}(r_\3,r_\2,r_\1)$
\footnote{
$H_p^{\bar\rho\rho}(f,g)=\sum_{\bar\imath,{\imath}\in\BI} (f(\imath'))^*
D_{\bar\imath{\imath}}(r_\3,r_\2,r_\1)g({\imath})$}
satisfy the invariance property \rf{crossing2}.
\end{conj}

\begin{figure}[t] 
\epsfxsize12cm
\centerline{\epsfbox{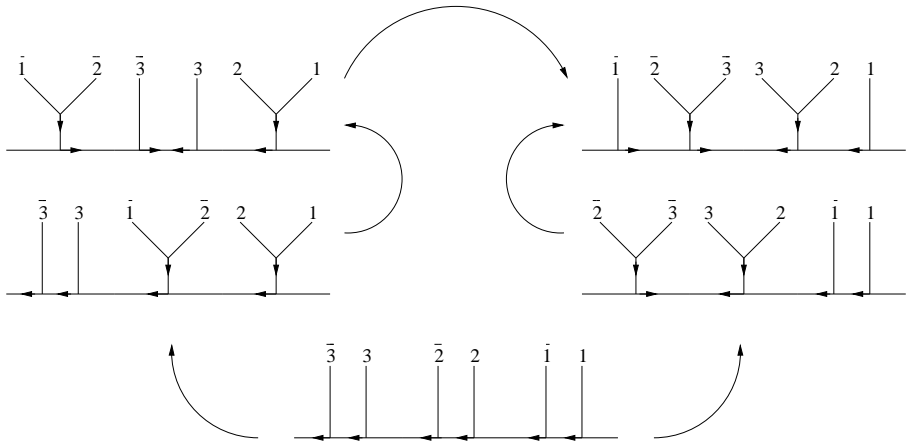}}
\caption{Proof of invariance under $\SF$.}
\label{scprodinvfig} \end{figure}


In other words, the data appearing in the
relationship \rf{spfus} characterize a 
hermitian form on spaces of conformal blocks 
canonically associated with any vertex algebra
$V$ that has the factorization property.

\subsubsection{}

Let us note that validity of the 
conjecture above is known in the case of rational CFT's.
Indeed, let us keep in mind that
according to \S \ref{H-3pt} above one may express the
three point function in the V-minimal model in terms
of the numbers $D_{\imath\bar{\imath}}(r_\3,r_\2,r_\1)$ introduced 
in the conjecture above. In the case that the operators $\SZ_p$, $\SB_p$
and $\SF_{pq}$ are represented by matrices it is easy to figure out 
an expression of $D_{\imath\bar{\imath}}(r_\3,r_\2,r_\1)$ in terms
of the matrix elements of $\SZ_p$ and $\SF_{pq}$. This expression 
coincides with the formula for the three point function that was 
obtained as a special case of the general formalism developed
in  \cite{FRS4} for the
description of correlation function in rational CFT. 
Invariance of the corresponding hermitian form
follows from the relations satisfied by the operators $\SZ_p$, $\SB_p$
$\SF_{pq}$, $\SS_p$ that were discussed in the previous section.

Our main point is of course to propose that a similar 
relationship also holds in nonrational cases. And indeed, given that
there exists a factorization of the form \rf{factmupl} it is not
hard to show
invariance under 
${\mathsf F}$ by considering the sequence of transformations
indicated in Figure \ref{scprodinvfig}. 
Invariance under ${\mathsf B}$ is verified similarly.
Invariance under ${\mathsf Z}$ follows from the invariance under
$\SF$
thanks to \rf{fustriv}. 
The conjecture is furthermore supported by the results from \cite{TL} 
reviewed in Section \ref{Liourev} above.

\subsection{Unitary fusion}

There is a generalization of the
tensor product for unitary representations 
of certain vertex algebras that has the 
virtue to make unitarity of the resulting representation 
manifest. The underlying theory is closely related to the 
theory of superselection sectors from algebraic
quantum field theory \cite{FRS,Ha}. 
We will in the following briefly discuss a reformulation 
called ``Connes-fusion'' \cite{Wa}.

In order to simplify the exposition, 
we will restrict attention to the case of the Virasoro algebra
with $c>25$. What follows is a sketch of the
picture that would result from using the results of 
\cite{TL} within a theory of 
``Connes-fusion'' of representations of 
${\rm Diff}(S_1)$ along the lines of \cite{Wa}. The author believes 
that similar things can be done for many other vertex algebras,
which would allow one to show that the canonical
hermitian form proposed in the previous subsection 
is positive definite in the case of unitary representations.

\subsubsection{}

It is temporarily useful to replace the states $v_m$ 
within the representations $R_m$, $m=1,2$ by the
vertex operators $\SV_m\equiv \SV_m(v_m)$ which generate the states 
$v_m$ from the vacuum, $v_m=\SV_m\Omega$. 
We want to think of representations $R_m$, $m=1,2$
as representations of ${\rm Diff}({\rm I}_m)$
associated to the 
intervals ${\rm I}_m$, respectively. 
The vertex operators $\SV_m$ should therefore commute with the action of
${\rm Diff}({\rm I}_m^c)$ as 
\begin{equation}\label{intertw}
\pi_{\sst R_m}^{}(g)\SV_m=\SV_m\pi_{\0}(g)\quad{\rm for\;\;all}\quad
g\in{\rm Diff}({\rm I}_m^c)\,,
\end{equation}
with $\pi_{\0}(g)$ being the action of  
${\rm Diff}({\rm I}_m^c)$ on the
vacuum representation $V$. Operators with such a property can be
constructed from the chiral vertex operators 
$\SY_{\al_\3\al_\1}^{\al_\2}({\mathfrak v}_\2|z)$ 
as
\begin{equation}\label{smearCVO}
\SV_m\,\equiv\,\int_{{\rm I}_m}dx \;f_m(x)\,
\SY_{r_m 0}^{r_m}({\mathfrak v}_m|e^{ix})\,,\qquad m=1,2\,,
\end{equation}
$f$ being a smooth function with support in ${\rm I}_m$. Operators 
like the one defined in \rf{smearCVO} will be unbounded in general,
but bounded operators can be obtained by taking the phase of its
polar decomposition.

Let us then
consider the spaces $\CV_m$ of {\it bounded} intertwiners $\SV_m:V\ra R_m$
which satisfy \rf{intertw}. On $\CV_\1\ot\CV_\2$ define an inner product
by 
\begin{equation}\label{connesfus}
\big\langle\,w_\1\ot w_\2\,,\,v_\1\ot v_\2\,\big\rangle\,=\,
\big\bra\,\Omega\,, \,\SW_\2^{*}\SV_\2^{}\,.
\SW^*_\1\SV^{}_\1\,\Omega\big\ket
\end{equation}
The Hilbert space completion of $\CV_\1\ot\CV_\2$ is denoted
$R_\1\boxtimes R_\2$. 
We observe that the scalar product of the 
``fused'' representation is defined in terms of the
conformal block $G^{\si_\1}_{\0\tau}\big[
\begin{smallmatrix} {r_\2} & {\bar{r}_\1}\\ {\bar{r}_\2} & r_\1
\end{smallmatrix}\big]$
that had previously appeared in the relations \rf{spfus}.

\subsubsection{}

In the case of the Virasoro algebra one may deduce the validity of relations
\rf{spfus} from \rf{LiouF} by analytically continuing
$\al_s$ to the value $\al_s=0$. This allows one to
write $ \lVert\,v_\1\ot v_\2\,\rVert^2$ in the form 
\begin{equation}\label{unieq}
\lVert\,v_\1\ot v_\2\,\rVert^2\,=\,
\int_\BU d\mu_{\2\1}(\al_s)\;
\lVert \, \SV^{\sst (2)}_{\al_s}(v_2,v_1)
\,\rVert^2_{R_{\al_s}}\,,
\end{equation}
where $\SV^{\sst (2)}_{\al_s}(v_\2,v_\1)$ 
is a certain vector in the {\it irreducible} 
representation $R_{\al_s}$ that may be written
as
\[
\SV^{\sst (2)}_{\al_s}(v_\2,v_\1)\,\equiv \,\SV_{s\1}^{\2} \,v_\1\,=\,
\int_{{\rm I}_\2}dx \;f_\2(x)\,
\SY_{\al_s \al_\1}^{\al_\2}({\mathfrak v}_\2|e^{ix})\,\SV_\1\,\Omega\,,
\] 
provided that $\SV_\2$ can be represented in the form \rf{smearCVO}.
Note that the space $R_\1\boxtimes R_\2$ is naturally a representation of 
${\rm Diff}({\rm I}_1)\times {\rm Diff}({\rm I}_2)$. 
Equation \rf{unieq} 
is interpreted as an expression for the unitary equivalence
\begin{equation} \label{unieq2} 
R_\1\,\boxtimes\,R_\2\,\simeq\,\int_\BU^{\oplus} d\mu_{\2\1}(s)\;R_{s}
\end{equation}
which implies in particular the fact that the representation of 
${\rm Diff}({\rm I}_1)\times {\rm Diff}({\rm I}_2)$ on
$R_\1\boxtimes R_\2$ can be extended to 
a representation of ${\rm Diff}(S^1)$. 
The factorization 
$
d\mu_{\2\1}(s)=d\mu_{\rm\sst Pl}(s) D(r_\3,r_\2,r_\1)$ 
then allows us to rewrite \rf{unieq2} as
\begin{equation} 
R_\1\,\boxtimes\,R_\2\,\simeq\,\int_\BU^{\oplus} d\mu_{\rm\sst Pl}(s)\;R_s\ot
{\rm Hom}(R_\1\boxtimes R_\2;R_s) \,,
\end{equation}
where ${\rm Hom}(R_\1\boxtimes R_\2;R_s)$ is the one-dimensional 
Hilbert space of intertwiners with metric given by
$D(r_\3,r_\2,r_\1)$.

\subsection{Associativity of unitary fusion}

It should be possible to show on
general grounds that the fusion operation is associative,
\begin{equation}\label{asso1}
(R_\1\boxtimes R_\2)\boxtimes R_\3\,\equiv\,R_\1\boxtimes 
R_\2\boxtimes R_\3\,\equiv\,
R_\1\boxtimes (R_\2\boxtimes R_\3)\,.
\end{equation}
Indeed, let us consider 
\begin{equation}\label{asso2}
\lVert\,(v_\1\ot v_\2)\ot v_\3\,\rVert^2 \quad{\rm and}\quad 
\lVert\,v_\1\ot (v_\2\ot v_\3)\,\rVert^2 
\end{equation}
The left hand side and the right hand side of \rf{asso2} can be 
represented respectively as 
\begin{align}\label{s-ch}
&\int_\BU d\mu_{\1\2}(\al_s)\;
\big\bra\,\Omega, \SV_\3^{*}\SV_\3^{}\,\SV^*_{\2\1}(\al_s)
\SV^{}_{\2\1}(\al_s)\,
\Omega\,
\big\ket\,,\qquad \SV^{}_{\2\1}(\al_s)\,\equiv\,\SV_{s\1}^{\2}\SV_\1^{}\,,\\
&\int_\BU d\mu_{\2\3}(\al_u)\;\big\bra\,\Omega, 
\SV_{\3\2}^{*}(\al_u)\SV_{\3\2}^{}(\al_u)\,
\SV^*_{\1}\SV^{}_{\1}\,
\Omega\,\big\ket\,,\qquad \SV^{}_{\3\2}(\al_u)\,\equiv\,
\SV_{u\2}^{\3}\SV_\2^{}\,.
\label{u-ch}\end{align}
It is useful to note that the compositions of chiral vertex operators
which appear in \rf{s-ch} and \rf{u-ch} correspond to the diagrams
on the left and right in the middle line of Figure \ref{scprodinvfig}, 
respectively. From this diagrammatic representation it is easily
seen that \rf{s-ch} and \rf{u-ch} are 
both equal to 
\begin{equation}
\bra\,\Omega, 
\SV_\3^{*}\SV_\3^{}\,\SV_\2^{*}\SV_\2^{}\,\SV^*_\1\SV^{}_\1\,
\Omega\,\ket\,\equiv\,\lVert\,v_\1\ot v_\2\ot v_\3\,\rVert^2\,,
\end{equation}
corresponding to the diagram on the bottom of Figure \ref{scprodinvfig},
which makes the associativity of the fusion operation manifest.
By using \rf{unieq} one may rewrite \rf{s-ch} and \rf{u-ch} 
respectively in the form
\begin{align}\label{s-ch+}
\int_\BU d\mu_{\1\2\3}(\al_\4)\;
\big\lVert\, & \SV_{\al_\4,s}^{\sst (3)}(v_\3,v_\2,v_\1) \,\big\rVert^{2}\,, 
\\[-1.5ex]
& \SV_{\al_\4,s}^{\sst (3)}(v_\3,v_\2,v_\1)
\,\equiv \,\int_\BU d\mu_{\1\2}(\al_s)\;
\SV_{\al_\4}^{\sst (2)}(\,v_\3\,,\,
\SV_{\al_s}^{\sst (2)}(v_\2,v_\1)\,) \nonumber \\
\int_\BU d\mu_{\1\2\3}(\al_\4)\;
\big\lVert\, & \SV_{\al_\4,u}^{\sst (3)}(v_\3,v_\2,v_\1) \,\big\rVert^{2}\,,  
\\[-1.5ex]
& \SV_{\al_\4,u}^{\sst (3)}(v_\3,v_\2,v_\1)
\,\equiv \,\int_\BU d\mu_{\2\3}(\al_u)\;
\SV_{\al_\4}^{\sst (2)}(\,
\SV_{\al_u}^{\sst (2)}(v_\3,v_\2)\,,\,v_1\,)\,. \nonumber
\end{align}
These relations may both be seen as expressions for the
unitary equivalences
\begin{align}
\label{fusion21}
(\,R_\1\,\boxtimes\,R_\2\,)\,\boxtimes\,R_\3\,
\simeq\,\int_\BU^{\oplus} d\mu_{\rm\sst Pl}(\al)\;R_\al\ot L^2(\BU,
d\mu_{(\1\2)3}^\al)\,,\\
\label{fusion32} R_\1\,\boxtimes(\,R_\2\,\boxtimes\,R_\3\,)
\simeq\,\int_\BU^{\oplus} d\mu_{\rm\sst Pl}(\al)\;R_\al\ot L^2(\BU,
d\mu_{\1(\2\3)}^\al)\,,
\end{align}
where 
\[
\begin{aligned}
& d\mu_{(\1\2)3}^{\al}(\al_s)=d\mu_{\rm\sst Pl}(\al_s)\;
D(\al_s,\al_\2,\al_\1)D(\al,\al_\3,\al_s)\\
&
d\mu_{\1(\2\3)}^\al(\al_u)=d\mu_{\rm\sst Pl}(\al_u)\;
D(\al_u,\al_\3,\al_\2) D(\al,\al_u,\al_\1)
\end{aligned}\]
It should be noted that 
the Hilbert spaces $L^2(\BU,d\mu_{(\1\2)3}^\al)$  and 
$L^2(\BU,d\mu_{\1(\2\3)}^\al)$
which appear in \rf{fusion21} and \rf{fusion32}, respectively,
are nothing but different models for Hilbert-subspaces of 
${\rm Hom}(R_\1\boxtimes R_\2\boxtimes R_\3;R_\al)$. 
These Hilbert spaces are canonically 
isomorphic to the spaces of conformal blocks $\CH_{\si_\1}$ and
$\CH_{\si_\2}$ associated to the markings on the left and right of 
Figure \ref{fmove}, respectively. 
It therefore follows from the associativity of the fusion product
that there exists a one-parameter family of 
unitary operators 
$\SF: \CH_{\si_\1}\ra \CH_{\si_\2}$ that 
represents the unitary equivalence between the 
representations \rf{fusion21} and \rf{fusion32}, respectively.

\subsection{Discussion}

The author believes that the link between the hermitian form on
spaces of conformal blocks and unitary fusion has
not received the attention it deserves.
More specifically, 
there are two reasons why the authors believes that the connection 
between the unitary fusion and the hermitian form on
spaces of conformal blocks is worth noting and being better
understood:

On the one hand, it offers an explanation for the positivity
of the coefficients $D_{\imath\bar\imath}(r_\3,r_\2,r_\1)$ 
defining the hermitian form $H_V$ in the unitary case, thereby elevating it to
a scalar product.

If, on the other hand, one was able to show on a priori grounds that
the representation ${\rm Diff}({\rm I}_1)\times {\rm Diff}({\rm I}_2)$ on
$R_\1\boxtimes R_\2$ can be extended to 
a representation of ${\rm Diff}(S^1)$ then one might use this 
as a basis for a  conceptual proof of the factorization property
\rf{genfusion} in genus 0 along the lines sketched above. 

It does not seem to be possible, however, to give a simple
explanation of the factorization \rf{factmupl}
in Conjecture \ref{hermformconj} from this 
point of view. This deep property seems to
require new ideas for its explanation. We see it as
a hint towards an even deeper level of understanding CFT
in its relation to the harmonic analysis of ${\rm Diff}(S_1)$, 
or some extension thereof. 

\section{Outlook}
\setcounter{equation}{0}

First we would like to stress that the class of nonrational CFT
covered by the formalism described in this paper is expected to
contain many CFTs of interest.
To illustrate this claim, let us formulate the following conjecture.

\subsection{Modular functors from W-algebras and Langlands duality}

We would finally like to formulate a conjecture. Let $W_k({\mathfrak g})$ 
be the W-algebra constructed in \cite{FF1,FF2}
\begin{conj} 
There exists a family of stable unitary modular functors
\[
\big(\,\Sigma\,,\,{\mathfrak g}\,,\,k\,\big)\,\longmapsto
\,\big({\mathfrak H}_{{\mathfrak g},k}(\Sigma)\,,\,
\SM_{{\mathfrak g},k}(\Sigma)\,\big)
\]
that is canonically isomorphic to either
\begin{quote}
the space of conformal blocks for certain classes of unitary 
representations of the W-algebra $\CW_k({\mathfrak g})$ 
with its natural unitary mapping class group action,
\end{quote}
or
\begin{quote}
the space of states obtained via the quantization of the 
higher Teichm\"uller spaces \cite{FG1,FG2} 
together with its canonical
mapping class group action
\end{quote}
such that Langlands duality holds: There is a canonical 
isomorphism
\[
\big({\mathfrak H}_{{\mathfrak g},k}(\Sigma)\,,\,
\SM_{{\mathfrak g},k}(\Sigma)\,\big)
\;\simeq\;
\big({\mathfrak H}_{{}^{\rm\sst L}\!{\mathfrak g},\check k}(\Sigma)\,,\,
\SM_{{}^{\rm\sst L}\!{\mathfrak g},\check k}(\Sigma)\,\big)\,,
\]
with ${}^{\rm\sst L}\!{\mathfrak g}$ being the Langlands dual to the
Lie algebra ${\mathfrak g}$ and $\check k$ being related 
to $k$ via $(k+h^{\vee})r^{\vee}=(\check k+h^{\vee})^{-1}$,
$h^{\vee}$ being the dual Coxeter number.
\end{conj}

\subsection{Boundary CFT}

It seems interesting to note a link to boundary CFT.
In the case of the $V$-minimal model one expects 
following Cardy's analysis \cite{Ca} to find
a one-to-one correspondence between conformal boundary 
conditions and irreducible representations. There 
should in particular exist a distinguished boundary 
condition $B_\0$ which corresponds to the vacuum representation.

This boundary condition is fully characterized by the 
measure appearing in the expansion of the corresponding
boundary state into the Ishibashi-states $|r\ket\!\ket$ which preserve
the full chiral algebra $V$,
\begin{equation}
|B_\0\ket\,=\,\int_\BU d\mu_{\rm B_\0}^{}(r)\;|r\ket\!\ket\,.
\end{equation}
It is not hard to see that the two-point function $\langle
V_\2(z_\2,\bz_\2)V_\1(z_\1,\bz_\1)\rangle_{B_\0}$
in the presence of a boundary with condition $B_\0$ 
is proportional to 
$G^{\si_\1}_{\0\tau}\big[
\begin{smallmatrix} {r_\2} & {\bar{r}_\1}\\ {\bar{r}_\2} & r_\1
\end{smallmatrix}\big]$. The expansion \rf{spfus} describes
the OPE of the two fields $V_\1$, $V_\2$. It easily follows
from these observations that the one-point function (in a suitable
normalization) coincides with the Plancherel-measure $d\mu_{\rm\sst Pl}(s)$,
\begin{equation}
d\mu_{\rm B_\0}^{}(r)\,=\,d\mu_{\rm\sst Pl}(r)\,.
\end{equation}
We take this observation as an intriguing hint concerning the generalization
of our considerations to boundary CFT.

\subsection{Nonrational Verlinde formula?}

In the case of rational CFT one can deduce a lot of useful 
relations \cite{MS,MS2} between the defining data of a modular functor 
from the relations \rf{zrel}-\rf{pentarel}, 
\rf{onetor:a}-\rf{twotorrel} and \rf{fustriv}-\rf{brtriv}. 
These relations give the key to some derivations of the famous 
Verlinde formula.
Much of this remains intact in the nonrational case,
as the example of formula \rf{Sformula} illustrates.

A fundamental difference comes from the fact that the
vacuum representation is {\it not} in the support of 
$\mu_{\rm\sst Pl}$. This implies that objects like
$\displaystyle
{\rm F}_{\al\be} \big[ {}_{\al_\4}^{\al_\3} {}_{\al_\1}^{\al_\2} \big]$
or $S_{\al\be}(\ga)$
are not necessarily well-defined at $\be=0$. 
This means that many of the relations valid in rational CFT
do not have obvious counterparts in the nonrational case.

As a particularly 
interesting example let us note that in the case of the
minimal models one has the relation \cite{R}
\begin{equation}
{\rm F}_{0r} \big[ {}_{r_\2}^{r_\2} {}_{r_\1}^{r_\1} \big]
{\rm F}_{r0} \big[ {}_{r_\1}^{r_\2} {}_{r_\1}^{r_\2} \big]
\,=\,\frac{S_{0r}S_{00}}{S_{0r_\2}S_{0r_\1}}\,,\qquad S_{r_\1r_\2}\equiv 
S_{r_\1r_\2}(0)\,.
\end{equation}
As explained above, the left hand side does not have an
obvious counterpart in the nonrational case in general.
However, in the case of the Virasoro algebra with $c>25$ 
it turns out that
\begin{equation}
{\rm F}_{\al 0}'' \big[ {}_{\al_\4}^{\al_\3} {}_{\al_\1}^{\al_\2} \big]\,=\,
\lim_{\be\ra 0}\,\be^2\,
{\rm F}_{\al\be} \big[ {}_{\al_\4}^{\al_\3} {}_{\al_\1}^{\al_\2} \big]
\end{equation}
exists and satisfies the relation
\begin{equation}\label{FFrel}
{\rm F}_{0\al}^{} \big[ {}_{\al_\2}^{\al_\2} {}_{\al_\1}^{\al_\1} \big]
{\rm F}_{\al0} ''\big[ {}_{\al_\1}^{\al_\2} {}_{\al_\1}^{\al_\2} \big]
\,=\,\frac{B_0\,B(\al)}{B(\al_\2)B(\al_\1)}\,,
\end{equation}
where $B(\al)=\sin 2b(\al-\de)\sin2b^{-1}(\de-\al)$. 
Equation \rf{FFrel} can be verified with the help of
the exlicit expressions for the objects involved. 

The  relation \rf{FFrel} is particularly suggestive
in view of the fact that $S_0{}^r$ gets identified with the
so-called quantum dimension 
in the correspondence between
modular functors and modular tensor categories \cite{BK1}.
What appears on the right hand side of \rf{FFrel} is
related to the measure $d\mu_{\rm\sst Pl}$ via
$d\mu_{\rm\sst Pl}(\al)=d\al \,B(\al)$, with $d\al$ being
the standard Lebesque measure on $\BT$.

This measure
can be seen as the most natural counterpart of the
quantum dimension in the nonrational case. This 
is seen most clearly when considering the quantum
group structure\footnote{More precisely: weak Hopf algebra structure} 
associated
to a rational modular functor \cite{Pf}. The quantum
dimension represents the weight of a representation 
in the Plancherel (or Peter-Weyl) 
decomposition of the space of functions
on the quantum group. As mentioned above, there is 
a quantum group ``dual'' to the modular functor 
defined by the representation theory of the 
Virasoro algebra with $c>25$ \cite{PT1,TR}. The natural
measure appearing in the  decomposition of the space of functions
on the corresponding quantum group is precisely $d\mu_{\rm\sst Pl}$ 
\cite{PT1}.   

It is clearly an important open task for the future 
to analyze  more systematically the set of relations that can be obtained 
in such a way.

\end{document}